\documentclass[preprintnumbers, floatfix,letterpaper, twocolumn,aps,prd,epsfig,nofootinbib,
]{revtex4-1}
\usepackage{bm,graphicx,dcolumn,epstopdf,epsf, latexsym,mathbbol, amssymb,amsmath,color,slashed, mathrsfs,mathcomp,simplewick}
\usepackage[center]{subfigure}
\usepackage{multirow}
\usepackage{physics}
\usepackage{makecell}
\usepackage[colorlinks,linkcolor=blue,citecolor=blue,urlcolor=blue]{hyperref}

\begin{document}
\allowdisplaybreaks
 \newcommand{\bq}{\begin{equation}} 
 \newcommand{\eq}{\end{equation}}
 \newcommand{\bqn}{\begin{eqnarray}}
 \newcommand{\eqn}{\end{eqnarray}}
 \newcommand{\nb}{\nonumber}
 \newcommand{\lb}{\label}
 \newcommand{\f}{\frac}
 \newcommand{\p}{\partial}
 \newcommand{\hong}{\textcolor[rgb]{0.85,0.42,0.00}}
\newcommand{\lan}{\textcolor[rgb]{0.2,0.2,0.7}}
\newcommand{\zi}{\textcolor[rgb]{0.51,0.21,0.67}}
\newcommand{\slan}{\textcolor[rgb]{0.40,0.40,0.78}}
\newcommand{\PRL}{Phys. Rev. Lett.}
\newcommand{\PLB}{Phys. Lett. B}
\newcommand{\PRD}{Phys. Rev. D}
\newcommand{\CQG}{Class. Quantum Grav.}
\newcommand{\JCAP}{J. Cosmol. Astropart. Phys.}
\newcommand{\JHEP}{J. High. Energy. Phys.}
\title{Pre-inflationary perturbations from closed algebra approach in loop quantum cosmology}
\author{Bao-Fei Li${}^{a, b}$}
\email{Bao-Fei$\_$Li@baylor.edu} 
\author{Tao Zhu${}^{a, b}$}
\email{zhut05@zjut.edu.cn} 
\author{Anzhong Wang${}^{a, b}$}
\email{anzhong$\_$wang@baylor.edu; Corresponding author} 
\author{Klaus Kirsten${}^{c}$}
\email{klaus$\_$kirsten@baylor.edu} 
\author{Gerald Cleaver${}^{d}$}
\email{gerald$\_$cleaver@baylor.edu} 
\author{Qin Sheng${}^{c}$}
\email{qin$\_$sheng@baylor.edu} 

\affiliation{${}^{a}$ Institute for Advanced Physics $\&$ Mathematics, Zhejiang University of Technology, Hangzhou, 310032, China
\\${}^{b}$ GCAP-CASPER, Physics Department, Baylor University, Waco, TX 76798-7316, USA
${}^{c}$ GCAP-CASPER, Mathematics Department, Baylor University, Waco, TX 76798-7328, USA\\
${}^{d}$ EUCOS-CASPER, Physics Department, Baylor University, Waco, TX 76798-7316, USA
}

\date{\today}

\begin{abstract}

In this paper, the scalar and tensor perturbations in the closed algebra approach of loop quantum cosmology are studied. Instead of the distant 
past in the contracting phase, we choose the moment at which the initial conditions are imposed to be the silent point, which circumvents the problem 
due to the signature change in the super-inflationary phase and results in a well-defined Cauchy problem. For the ultraviolet and infrared modes, 
different approaches are applied in order to obtain analytical solutions with high accuracy. While previous numerical simulations reveal an exponentially 
divergent power spectrum in the ultraviolet regime, when the initial conditions are imposed in the remote contracting phase, we find a special set 
of initial conditions at the silent point, which can  reproduce results that are consistent with current observations.

\end{abstract}

\maketitle

\section{Introduction}

As a non-perturbative and background-independent approach to quantizing general relativity, loop quantum gravity (LQG) presents a basic framework in which discrete structures of some fundamental geometric quantities, such as areas and volumes, are found at the quantum level \cite{ds2010, rv, rs1995}. However, as LQG is concerned with physics at the Planck scale,  it's almost impossible to  directly detect the predictions from LQG  by any man-made terrestrial experiments  in the near future. This makes primordial cosmology an invaluable arena for investigating various aspects of quantum gravity, including testing LQG. In the cosmological settings, due to the homogeneity and isotropy of the Universe, the quantization of cosmology can be studied by  a symmetry reduced version of LQG \cite{abl2003, as2011}. Although different quantization approaches  result in different models and thus  distinctive phenomenologies \cite{ydm09, dl2018, lsw2018-1}, in this paper,  we will mainly focus on  loop quantum cosmology (LQC),  which is one of the most developed quantum cosmological theories.

In LQC, the classical singularity in the standard big-bang model is replaced by the quantum bounce due to the pure quantum geometric effects at the Planck scale \cite{a2009,acs2008,s2009}. The pre-inflationary dynamics of LQC has been studied extensively in the literature \cite{ad2011,lb2013,ssww2017,ssw2018,ssqz2018,bcl2018, zhu_universal_2017, zwcks2017, jin_preinflationary_2018,lsw2018-2,zhu_primordial_2018,wu_nonadiabatic_2018,mbs2017}.  {Basically,   the background evolution in LQC is symmetric with respect to the bounce,  and the slow-roll inflation turns out be an attractor in a variety of inflationary potentials when the Universe is sourced by a single scalar field \cite{zhu_universal_2017, zwcks2017, lsw2018-2, jin_preinflationary_2018,bbmm}. One of the most important unknowns regarding the dynamics of the background in LQC is the way to set initial conditions. In principle, the initial conditions can be set either in the remote past of the contracting phase \cite{lb2013} or at the bounce point when the critical energy density is attained \cite{aan2013-2}. The most immediate physical consequence resulted from the choice of different initial conditions lies in the duration of the inflationary phase.  For example, if the initial conditions are chosen in the prebounce phase,  the preferred number of  inflationary e-folds  is around 140 for the chaotic inflation \cite{lb2013, zhu_primordial_2018, zhu_universal_2017, zwcks2017, wu_nonadiabatic_2018}. Whereas,  if the initial conditions are set at the bounce point, the most probable e-folds number can be as large as $10^{12}$ \cite{mbs2017}.  Even though, there seems huge differences in  the number of inflationary e-folds when different ansatzes for setting initial conditions are employed, these differences are not substantial in the sense that the most probable e-folds number in one ansatz can be the fine-tuned one in the other ansatz. On the other hand, more important differences show up from different approaches to cosmological perturbations.    }{

Up to now, there are  at least three different approaches to study the primordial power spectra in LQC,  the dressed metric \cite{aan2012, aan2013-1, aan2013-2} \footnote{The hybrid approach \cite{fmmmo2012,fmmmo2013,gmmo2014,gbm2016} is quite similar to the dressed metric one, and the obtained results are observationally in-distinguishable \cite{do2016}, so in this paper we do not consider it separately.}, closed algebra \cite{bhks2008, cmbg2012, cbgv2012} and separate universes \cite{we2016, we2017}. In the dressed metric approach, the background metric is quantized by the loop approach while its perturbative degrees of freedom are quantized following the Fock quantization procedures. As long as the energy density in the perturbations remain small as compared to the Planck energy, the quantum dynamics of the perturbations can be described by a quantum field propagating on a dressed background spacetime. On the other hand, the closed algebra approach requires no knowledge of underlying quantum theory. It's based on the effective constraints arising from the quantum corrections (either inverse triad or holonomy connections) and the requirement that effective constraint algebra be closed after the quantum corrections are taken into account. This anomaly-free condition helps fix the counter terms in the effective constraint and the resulting Poisson bracket of the scalar constraint with itself is deformed  by a   factor $\Omega (=1-2\frac{\rho}{\rho_c})$,  as compared to the classical case.  The same  factor is also present in the quantum equations of motion for cosmological perturbations and lies at the center of the problem called signature change  \cite{m2014, bm2015} \footnote{Actually the existence of the signature change near the LQC bounce is still an open question, for more details, see the discussions in \cite{we2017} and references therein.}. Finally, in the separate Universe approach, the cosmological spacetime with small perturbations is discretized into a lattice and loop quantization is applied to each cell which is assumed to be homogeneous and non-interacting with each other. In this way, the dynamics of cosmological perturbations can be approximated by the effective equations when the wave functions in each cell is sharply peaked. Although up to now only the scalar perturbations in the longitudinal gauge is quantized and the results are applicable only to the IR modes, it's the only one that performs loop  quantization to both background  and perturbations at the same time. 

In the following, we will mainly focus on the closed algebra approach. Previous work \cite{lcbg2013,bbgs2016,bgsl2015,g2016} reveals the generic behavior of primordial power spectra in this approach, namely, scale invariance in the IR regime, oscillations for the intermediate scales and an exponential growth in the UV regime\footnote{In the dressed metric  approach, the power spectra are enhanced and oscillating for the intermediate scales and scale-invariant in the UV regime \cite{aan2013-2,am2015}. }. In particular, the exponential growth in the UV regime is due to the fact that the initial conditions are imposed in the contracting phase while the equations of motion for the cosmological perturbations become elliptic near the bounce point \cite{lcbg2013}. The divergence of power spectra in the UV regime is thought to be inconsistent with the observations as the observable window of the wave numbers in CMB is most probably located in this regime, considering the preferred number of inflationary e-folds is around 140 in LQC \cite{bbgs2016}.  So far, there exist mainly two proposals to modify the UV behavior in this approach. The first one  is  to impose the initial conditions  at the silent point so that the question of evolution of the mode function through the super-inflationary phase is avoided \cite{mlb2017}. The second proposal is to address the trans-Planckian problem via the modified dispersion relations (MDRs) \cite{mbg2018}. However, as the minimal non-zero length in LQG is still unknown due to the complexity of the length operator, MDRs should be regarded as a  phenomenological attempt towards solving the trans-Planckian problem in LQC. 

In this paper, we will simply follow the first approach and choose the moment at which  the initial conditions are imposed to be the silent point,  so that the complications arising from the signature change near the bounce is  circumvented.  However, instead of imposing the initial conditions at the silent point, we flap the argument by asking what  conditions are required at this moment, in order to produce  predictions that are consistent with observations? To answer this question,   we first present our  analytical approximate solutions  of the mode functions in the UV regime \footnote{Recently, it was pointed out that the closed algebra approach might be applicable only to the infrared  modes \cite{we2017,mbg2018}.}, by applying the well-developed method, {\em the uniform asymptotic approximation (UAA) method},  which we have shown is very powerful in studying cosmological perturbations for various inflation models \cite{zhu_constructing_2014, zwcks2013, zwcks2014, zwcks2016, zwcks2017, zhu_power_2014,zhu_scalar_2015, zhu_field_2018} \footnote{All of the previous calculations of the power spectra in the UV regime rely on the numerical simulations.}. It is these analytical solutions that enable us to impose the observationally-consistent conditions at the end of inflation, and then evolve the mode functions backward until the silent point, whereby we read off the initial conditions at this silent point. Of course, this only tells us that if the initial conditions are chosen (at this silent point) in such a way, the resulted power spectra will be consistent with current cosmological observations. Clearly, the next question is what is the physics that leads to such conditions. Currently, we do not have a definite answer to it, and it is still under our investigations.  

The paper is organized as follows. Sec. II is devoted to a concise review on the background evolution of the LQC Universe  when the bounce is dominated by the kinetic energy of the scalar field. The pre-inflationary evolution of the scale factor and the scalar field is explicitly described by their analytical solutions of the Friedmann and Klein-Gordon equations. In Sec. III,  with the UAA method,  analytical solutions of the scalar mode function in the UV regime are derived. With the help of these analytical solutions, corrections to primordial power spectra due to the quantum gravitational effects at the Planck scale can now be parametrized in terms of the initial conditions imposed at the silent poin. We then focus on finding a particular set of initial conditions that can produce results consistent with the observations.  In this section, we also consider the IR regime\cite{bgsl2015}. In Sec. IV, similar analysis is carried out for the tensor perturbations. In particular, we first present the analytical solutions of the tensor modes in both UV and IR regimes, and then  find the relation between the initial conditions at the silent point and the quantum corrections to the power spectra.  Again, by requiring that the power spectrum at the end of inflation be the same as that obtained in general relativity, we are able to identify a set of initial condition at the silent point. Finally, we summarize our main results and give some comments in Sec. V.

Throughout the paper,  the Planck units are used in which $c=\hbar=G=1$ so that the Planck length, time and mass are all equal to unity. 

\section{The Pre-inflationary Evolution in Loop Quantum Cosmology}
\renewcommand{\theequation}{2.\arabic{equation}}\setcounter{equation}{0}

This section is mainly devoted to a review on the pre-inflationary evolution of the Universe in the spatially flat Friedmann-Lemaitre-Robertson-Walker (FLRW) background described by
\bqn
ds^2=-dt^2 +a(t)dx_i dx^i,
\eqn
where $a(t)$ is the cosmological scale factor and $t$  is cosmic time. In LQC, the effective dynamics of a flat FLRW Universe is governed by the modified Friedmann equation \cite{s2006, aps2006}
\bqn
\lb{friedmann}
H^2=\frac{8 \pi}{3m_{\text{Pl}}^2}\rho\left(1-\frac{\rho}{\rho_\text{c}}\right),
\eqn
where $H\equiv \dot a/a$ denotes the Hubble parameter and dot represents the derivative with respect to the cosmic time, $m^2_{\text{Pl}}=1/G$ is the Planck mass, $\rho$ is the energy density of the scalar field, and $\rho_\text{c}$ is the critical energy density which represents the maximum value of the energy density in LQC and can be approximated by $\rho_\text{c} \simeq 0.41 m_\text{Pl}^4$ \cite{m2004}. In this paper, we only consider inflation sourced by a single scalar field. Therefore, in the matter sector,  the effective equation of motion of the inflaton field  with a potential $V(\phi)$ is just  the Klein-Gordon equation given by 
\bqn
\lb{klein-gordon}
\ddot \phi +3 H \dot \phi +V_{,\phi}=0,
\eqn
where $V_{,\phi}\equiv dV(\phi)/d\phi$. As usual,  the property of the scalar field can be well described by its effective equation of state $\omega_\phi$, which is  defined via
\bq
\lb{eos}
\omega_\phi\equiv \frac{P}{\rho}=\frac{\frac{1}{2}\dot \phi^2-V}{\frac{1}{2}\dot \phi^2+V},
\eq
where $P$  represents the pressure  of the scalar field.

A robust prediction of LQC is the occurrence of a non-singular quantum bounce, which removes the initial singularity in the deep Planckian regime  purely due to the quantum geometric effects \cite{acs2008}. Eq.~(\ref{friedmann}) shows that the quantum bounce occurs at $\rho=\rho_\text{c}$, where the energy density reaches its maximum value and the Hubble parameter becomes zero. The background evolution with a bouncing phase has been extensively studied \cite{ad2011,lb2013,ssww2017,ssw2018}, and one of the main results is that, right following the quantum bounce, a desired slow-roll inflation phase is attractive.

In the cosmological settings, it is also convenient to introduce the conformal time $\eta$  via
\bqn
\lb{conformal}
\eta=\int_{t_\text{end}}^{t} \frac{dt'}{a(t')},
\eqn
so that at the end of the inflation $t=t_{\text{end}}$ and at the bounce time $t_\text{B}=0$, the corresponding conformal times are given, respectively, by 
\bqn
\eta_{\text{end}}=0,\;\;\;\;\;\eta_{\text{B}}=\int^{t_\text{B}}_{t_{\text{end}}}\frac{dt'}{a(t')}<0.
\eqn

Moreover, previous studies on the background evolution in LQC with a chaotic potential also show that if the initial conditions are imposed in the remote past of the contracting phase, the most probable state at the bounce is dominated by the kinetic energy \cite{lb2013}. Furthermore, for the kinetic-energy-dominated bounce, a universal description of the background in the bouncing phase, irrespective of the specific form of the potential, is also known\cite{zwcks2017}. Thus, in this paper, we will concentrate on the background when the quantum bounce is dominated by the kinetic energy of the scalar field, in which it has been shown  that the evolution of the universe between the quantum bounce and reheating can be universally divided into three different phases:  {\em bouncing, transition and  slow-roll inflation} \cite{zwcks2017,ssww2017,zhu_universal_2017,ssw2018,ssqz2018}.
In the rest part of this section, we will summarize the analytical solutions of the background evolution in these three phases.

\subsection{The Bouncing Phase}

The bouncing phase is characterized by the condition $\omega_\phi\approx 1$ for the kinetic dominated bounce. Therefore, the potential term in the evolution equations (\ref{friedmann}) and (\ref{klein-gordon}) can be ignored in the   leading-order approximation. This results in two simplified equations
\bqn
&&\lb{fri}H^2=\frac{4\pi G \dot \phi^2 }{3} \left(1-\frac{\dot \phi^2}{2\rho_{\text{c}}}\right),\\
&&\lb{kg}\ddot \phi+3 H \dot \phi=0,
\eqn
where we have ignored the potential terms as $\dot \phi^2 \gg V(\phi)$ near the bounce. Then, from the Klein-Gordon equation (\ref{kg}), it's straightforward to obtain
\bqn
\lb{phi_dot}
\dot \phi(t)= \pm \sqrt{2\rho_{\text{c}}} \left(\frac{a(t)}{a_\text{B}}\right)^{-3}.
\eqn
Here $\pm$ correspond to the cases that $\dot \phi_\text{B}$ is positive or negative at the bounce, respectively. Substituting this into Eq. (\ref{fri}), we   find
\bqn\lb{scalar_analytical}
a(t)=a_\text{B}\left(1+ \frac{t^2}{\tau^2_B}\right)^{1/6},
\eqn
where $\tau_B=t_{\text{Pl}}/\sqrt{\gamma_\text{B}}$ and $\gamma_\text{B}\equiv \frac{24\pi \rho_{\text{c}}}{m_{\text{Pl}}^4}\approx30.9$ is a dimensionless constant. From its definition in Eq.~(\ref{conformal}),
 the conformal time $\eta$ can be expressed in terms of the cosmic time as
\bqn
\eta(t)-\eta_B=t \,_2F_1\left(\frac{1}{6},\frac{1}{2},\frac{3}{2},- \frac{t^2}{\tau_B^2}\right),
\eqn
where $_2F_1(a,b,c,z)$ is the hypergeometric function. Also using the analytical solution of $a(t)$ in Eq. (\ref{scalar_analytical}), one   finds  that
\bqn \lb{phi_sol}
\phi(t)=\phi_{\text{B}} \pm \frac{m_{\text{Pl}}}{2\sqrt{3\pi}}\text{arcsinh}{\left(\frac{t}{\tau_B}\right)},
\eqn
and
\bqn \lb{phidot_sol}
\dot \phi(t)= \pm \frac{\sqrt{2\rho_\text{c}}}{(1+ t^2/\tau_B^2)^{1/2}}.
\eqn

\subsection{The Transition Phase}

This phase is characterized by a drastic drop of the equation of state $\omega_\phi$ from positive to negative unity. Two particular moments are generally used to characterize this phase. The first one is denoted by $t_c$ when the kinetic energy equals the potential energy, and at this moment, $\omega_{\phi}=0$. The second is $t_i$ at which $\omega_{\phi}$ drops to negative one third which signifies the beginning of the acceleration of the Universe and thus can also be 
regarded as the beginning of the slow-roll inflation.\footnote{To be precise, the  acceleration of the Universe  is not necessarily equivalent to the slow-roll inflation. However, 
once the Universe starts to accelerate, it quickly enters into the slow-roll inflation.} As the transition phase is shortly-lived, lasting only less than one e-fold, it is quite reasonable to assume that the analytical solutions (\ref{scalar_analytical}) and (\ref{phi_sol}) from the bouncing phase remain valid until $t_c$.  The validity of this assumption is checked in details in \cite{zwcks2017} with the chaotic and Starobinsky potentials, and the relative errors between the numerical and analytical solutions are shown to be less than $4\%$. Therefore, in this paper  we directly employ (\ref{scalar_analytical}) and (\ref{phi_sol}) for the analytical approximations of the scale factor and scalar field during this phase, respectively.

\subsection{The Inflationary Phase}

During the slow-roll inflationary phase, the Universe is dominated by the potential energy of the scalar field and therefore experiencing an exponential expansion. The evolution of the scale factor can thus be approximated by 
\bq
a(t) \approx a_i e^{H_{\text{inf}}\left( t-t_i \right)},
\eq
here $H_{\text{inf}}$ denotes the Hubble parameter in the sow-roll inflation and $a_i$ is the scale factor at the onset of the slow-roll inflation whose explicit expression depends on the form of the potential. Besides, the evolution of the scalar field is governed by the Klein-Gordon equation 
\bq
3H\dot \phi+\frac{d V(\phi)}{d \phi}\approx0.
\eq
Once the form of the potential is known, the solution $\phi$ can be found explicitly.

\section{The Primordial Scalar Perturbations}
\renewcommand{\theequation}{3.\arabic{equation}}\setcounter{equation}{0}

As the background evolution is known, it's appropriate  to proceed to cosmological perturbations in LQC.  In this section, we will focus on the scalar perturbations. First, the analytical approximations of the scalar mode functions will be given in both UV and IR regimes. Then, the relation between the initial conditions at the silent point and the power spectra in the UV regime will  be fixed by matching the analytical solutions in the bouncing, transition and inflationary phases. Finally, we will discuss the possibility of choosing particular initial conditions that can produce primordial power spectra that are consistent with current observations.  

\subsection{The Mode Function and Co-moving Hubble Horizon}

In the framework of the closed algebra approach, the mode function $\mu_S(\eta)$ for scalar perturbations obeys the equation \cite{bhks2008, cmbg2012, cbgv2012}
\bqn
\lb{scalarfun}
\mu_S''(\eta)+ \omega^2_k \mu_S(\eta)=0,
\eqn
where
\bqn
\lb{scalar1}
\omega^2_k&\equiv&\Omega(\eta) k^2-\frac{z_S''(\eta)}{z_S(\eta)},\\
\Omega(\eta)&\equiv& 1- \frac{2\rho(t)}{\rho_{\rm c}},\\
z_S(\eta) &\equiv& a \frac{\phi'}{\mathcal{H}} = a \frac{\dot \phi}{H},
\eqn
here $k$ is the co-moving wavenumber,  $\mathcal{H} \equiv a'/a$ and a prime denotes differentiation with respect to the conformal time $\eta$.

During  the  bouncing and transition phases, using the analytical approximations, namely, Eqs.~(\ref{scalar_analytical}) and  (\ref{phi_sol}), we find
\bqn\lb{omega}
\Omega(\eta) = \frac{\tau^2-1}{\tau^2+1},
\eqn
and
\bq
\frac{z_S''}{z_S}=\frac{\gamma_B\left(18+21\tau^2-\tau^4\right)}{9\tau^2 \left(\tau^2 +1\right)^{5/3}},
\eq
with $\tau\equiv t/\tau_B=t\sqrt{\gamma_B}$. In general, the qualitative evolution of the mode function $\mu_S$ depends on the sign of $\omega^2_k$. The modes with negative $\omega^2_k$ are  called decaying/growing modes, and these modes are outside the Hubble horizon. The modes with positive $\omega^2_k$ are called oscillatory modes and they are inside the Hubble horizon. The complications in the closed algebra approach lie in the $\Omega$ factor of Eq. (\ref{scalar1}) which will change signs at  $\rho=\frac{\rho_c}{2}$.  The moment where $\Omega(\rho = \rho_c/2) =0$ is called the silent point at which all the space points are de-correlated as the space-dependent term  drops out  from   Eq. (\ref{scalarfun}) and the two point functions on this surface become zero \cite{mlb2017}. In the expanding phase, the silent point  is always located at the end of the super-inflationary phase when $t=t_S\approx0.18$. According to the signature of $\Omega$, there are two distinctive regions:  the Euclidean regime where $\Omega$ is negative and $\frac{\rho_c}{2}<\rho\le \rho_c$; the Lorentzian regime where $\Omega$ is positive and $\rho<\frac{\rho_c}{2}$. 

Taking into account the sign change of $\Omega$ across the silent point,  the co-moving Hubble horizon (we will suppress the term ``co-moving" in the following for simplicity) can be defined by 
\bq
\lb{scalar_horizon}
\lambda^2_S=  \Omega \frac{ z_S}{z_S^{''}},
\eq
whose behavior in the post bounce phase is depicted in Fig. \ref{fig1}. At the  bounce, the Hubble horizon equals zero as $z_S=0$  at $t=t_B$. It remains negative during the super-inflationary phase from $t_B$ to $t_S$.  After $t_S$, the Hubble horizon continues to increase until $t=t_H$ when $z^{''}_S$ equals zero and the Hubble horizon becomes infinitely large. In the interval $ t_H<t< t_i$, the Hubble horizon is always negative which indicates all the relevant modes are now inside the horizon. Finally, $t_i$ represents the onset of slow-roll inflationary phase in which the Hubble horizon decreases monotonically. The black dot-dashed horizontal line in Fig. \ref{fig1} denotes a particular mode with  co-moving number $k$. It's outside the horizon at the bounce point as $\Omega$ is negative near the bounce. At time $t=t_k$, this particular mode enters into the horizon and it remains inside the horizon until $t_*$ when horizon crossing takes place during the slow-roll inflation.  Although $t_k$ and $t_*$  do depend on the wavenumber of the particular mode,  the generic behavior of scalar perturbations in the closed algebra is that all the modes experience horizon crossing  twice in the post bounce phase. The first time is at point $A$ indicating horizon-entry, and the second time is at point $B$ indicating horizon-exit.  As all the modes are outside the horizon at the bounce, the quantum effects near the bounce would affect all the modes from UV to IR regimes. This is different from the dressed metric approach in which there is a maximum wavenumber $k_B$, and the modes with wavenumber $k>k_B$ would not be affected by the dynamics around the bounce. 
 
\begin{figure}%
{
\includegraphics[width=8.5cm]{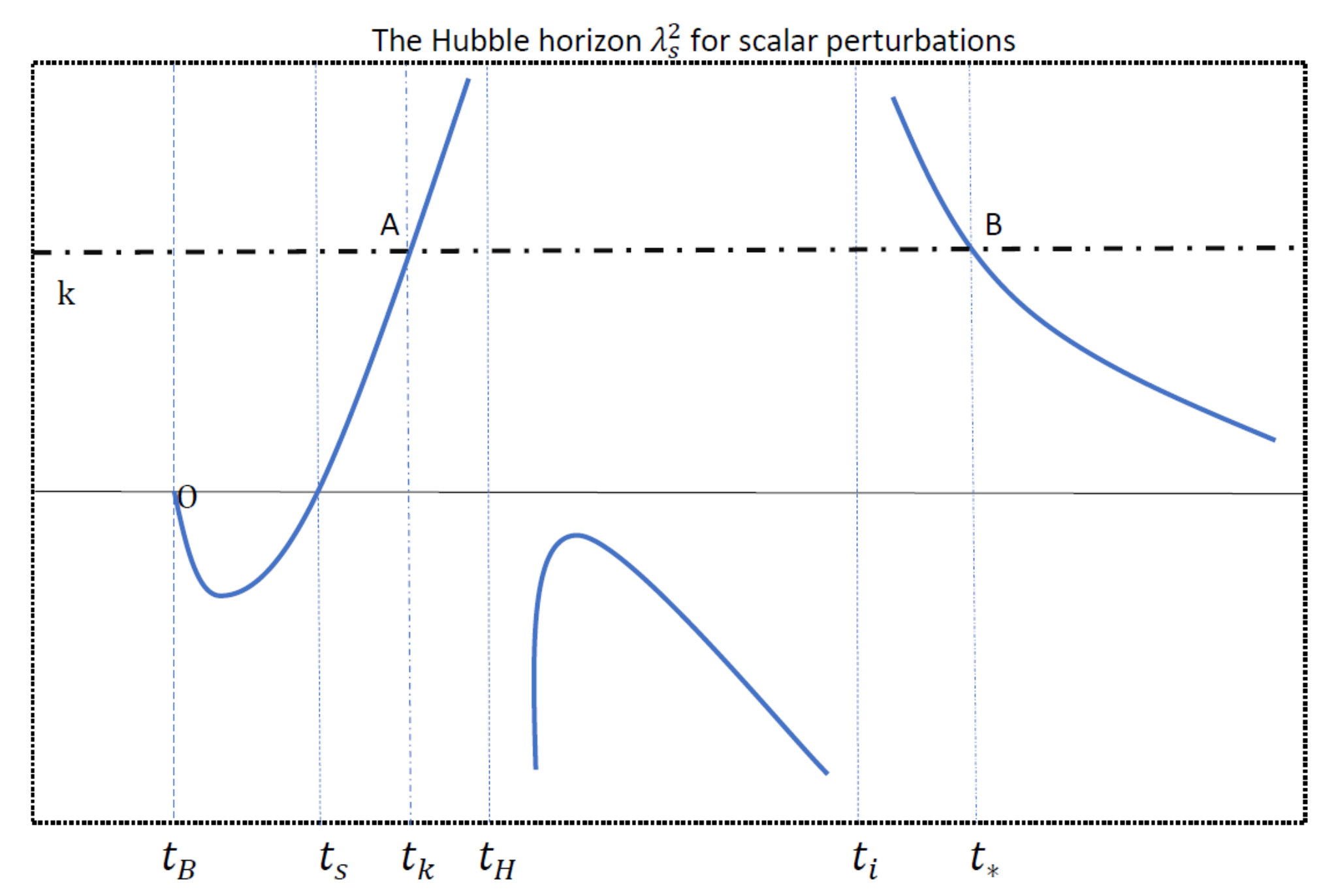}}
\caption{A schematic plot of the Hubble horizon defined in Eq. (\ref{scalar_horizon}).The solid curves show the evolution of the Hubble horizon $\lambda^2_S$ in the post bounce phase. $t_B$ denotes the time at the bounce, $t_S$ marks the moment when the super-inflationary phase is ended,  $t_H\approx 0.84t_{\text{Pl}}$ is the time when $z^{''}_S=0$ and $t_i$ is the time when the slow-roll inflation starts. The dot-dashed black line corresponds to the mode with co-moving wavenumber $k$ which enters  the horizon at $t_k$ (point $A$) and later exits the horizon at $t_*$ (point $B$) in the inflationary phase. }
\label{fig1}
\end{figure}

\subsection{Uniform Asymptotic Approximation and Analytical Approximations of the Mode Function}
 
In this subsection, we will  give an outline of  UAA in order to solve the mode function  equation (\ref{scalarfun}) (see \cite{zwcks2014} for a detailed exposition of the method for equations with one single turning point.).  First,  we need to put the mode equation  into the standard form
\bqn
\lb{3.1}
\frac{d^2\mu_S(y)}{dy^2}=\Big\{ g_S(y) + q(y) \Big\}\mu_S(y),
\eqn
where $y=-k\eta$ and 
\bqn
\lb{3.2}
g_S(y) + q(y) &=& \frac{1-\tau^2}{1+\tau^2}\nb\\
&+&\frac{\gamma_B\left(18+21\tau^2-\tau^4\right)}{9k^2\tau^2 \left(\tau^2 +1\right)^{5/3}}.
\eqn

At this point,  use can be made of the Liouville transformation, which is 
\bq
\lb{LT1}
U(\xi)=\chi^{\frac{1}{4}}\mu_S(y)  \quad \quad \text{and} \quad \quad  \xi'^2=\frac{|g_S|}{f^{(1)}(\xi)^2},
\eq
where $\chi \equiv \xi'^2$, $\chi'=d\chi/dy$ and 
\bq
\lb{LT2}
f(\xi)= \int^y \sqrt{|g_S(y)|}dy, \quad \quad f^{(1)}(\xi)^2=\frac{df(\xi)}{d\xi}.
\eq
Consequently, the mode function equation (\ref{scalarfun}) can now be transformed into 
\bq
\lb{scalarfun2}
\frac{d^2U}{d\xi^2}=[\pm f^{(1)}(\xi)^2+\psi(\xi)]U,
\eq
where the $\pm$ signs correspond to $g_S>0$ and $g_S<0$, respectively, and 
\bq
\psi(\xi)=\frac{q_t}{\chi}-\chi^{-3/4}\frac{d^2(\chi^{-1/4})}{dy^2}.
\eq
The key point in  UAA is to   {find a proper  choice} of $q(y)$ and $f^{(1)}(\xi)^2$ so that the errors can be made as small as possible. To be concrete, in order to fix $q(y)$, one needs to first note that $g_S(y) + q(y) $ has a second-order pole at $t=0$ (or equivalently $\eta=\eta_{\rm B}$), that is
\bqn
g_S(y) + q(y) &\to& \frac{2}{k^2 t^2}  \quad \text{as} \quad  t \to 0^+.
\eqn
According to the analysis of the error control function of UAA near the pole \cite{zwcks2016}, one has to choose 
\bqn
q(y) = - \frac{1}{4 k^2 t^2},
\eqn
for the convergence of error. So we have
\bqn
\lb{2.12}
g_S(y) &=&\frac{1-\tau^2}{1+\tau^2}+ \frac{\gamma_{\rm B} }{4k^2\tau^2}\nb\\
&&+\frac{\gamma_B\left(18+21\tau^2-\tau^4\right)}{9k^2\tau^2 \left(\tau^2 +1\right)^{5/3}}.
\eqn
Thus, for an arbitrary mode (any fixed k), in the expanding phase, $g_S$ is always a monotonically decreasing function of cosmic time as depicted in Fig. \ref{fig2}. Although the precise location of the turning point $t_+$, namely $g_S(t_+)=0$, does depend on the co-moving wavenumber $k$, qualitative behavior of $g_S$ remains the same for all the modes, that is, $g_S$ has only one turning point in the bouncing and transition phases for any given $k$.

\begin{figure}
{
\includegraphics[width=8cm]{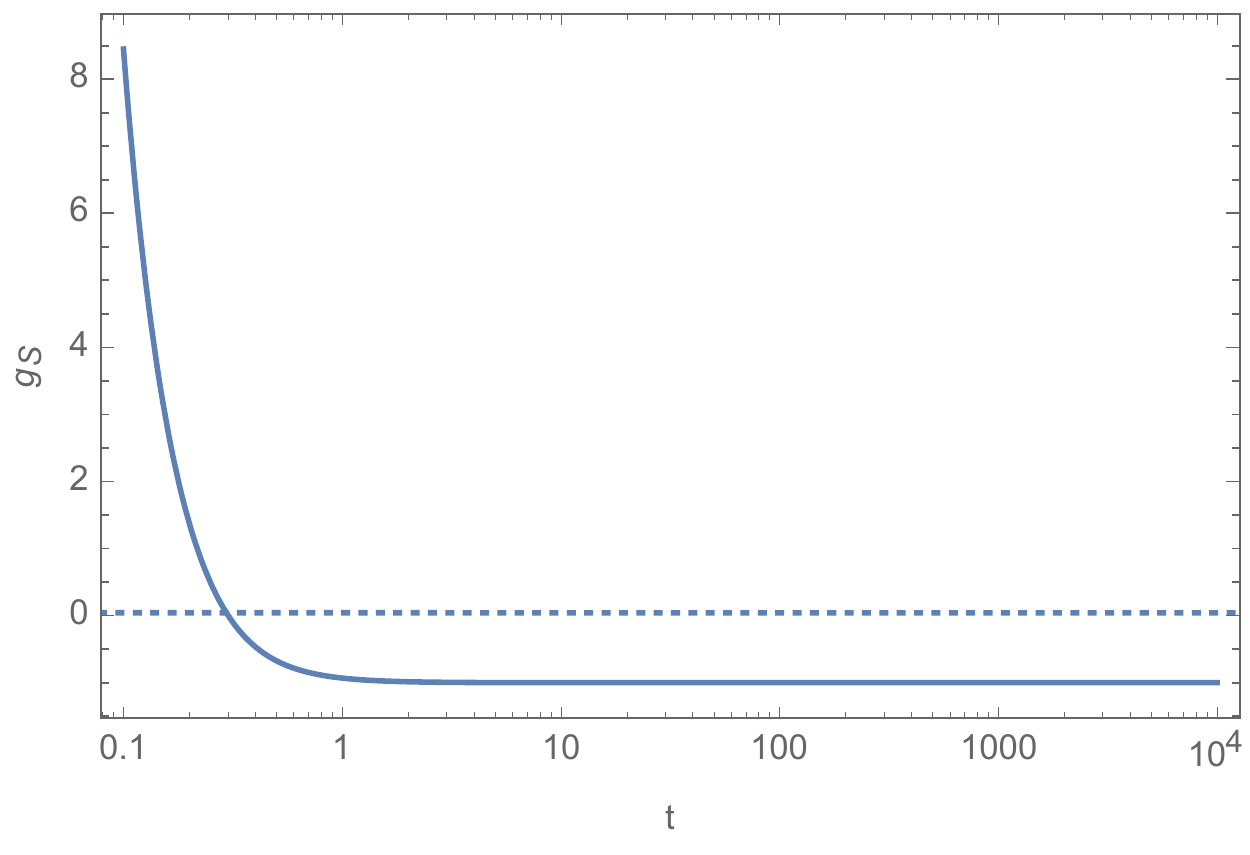}
}
\caption{This figure   {illustrates the general behavior of the function $g_S$. It shows explicitly that in the expanding phase when $t>0$, $g_S$ has only one turning point and it converges  quickly to negative unity at large times. When plotting it, we   had chosen $k=5$.}}
\label{fig2}
\end{figure}

Once the property of $g_S$ is known, we can make a good choice of $ f^{(1)}(\xi)^2$. Generally,  the choice of $ f^{(1)}(\xi)^2$ relies on the properties of the turning points of $g_S$ which, as discussed above, has only one single turning point in the expanding phase. As a result,  one can simply choose 
\bq
f^{(1)}(\xi)^2=\pm \xi,
\eq
again the $\pm$ signs are assigned, respectively,  in the regions where $g_S$ is positive/negative. Now that $q(y)$ and $ f^{(1)}(\xi)^2$ are fixed, the leading-order approximation of the mode function can be derived by simply ignoring the $\psi(\xi)$ term in Eq. (\ref{scalarfun2}),  {which}  leads to 
\bqn
\lb{3.9}
\mu_S(t) = \left(\frac{\xi_S}{ g_S}\right)^{1/4} \Big\{a_k \rm{Ai}\left(\xi_S\right) +b_k\rm{Bi}\left(\xi_S\right)\Big\},
\eqn
where $a_k$, $b_k$ are two integration constants, $A_i$ and $B_i$ are the Airy functions of the first and second kind, respectively. Besides, $\xi_S$ is given explicitly by the integral 
\bqn
\lb{3.10}
\xi_S= 
\begin{cases}
 \left(- \frac{3k}{2} \int_{t_{+}}^t \frac{\sqrt{g_S}}{a(t)} dt \right)^{2/3} &, t< t_+ \\
- \left( \frac{3k}{2} \int_{t_{+}}^t \frac{\sqrt{- g_S}}{a(t)} dt \right)^{2/3} &, t> t_+ .
\end{cases}
\eqn
Moreover, the mode function $\mu_S(t)$ satisfy the Wronskian condition 
\bq
\lb{3.11}
\mu_S \dot \mu_S^*-\mu_S^*\dot \mu_S=\frac{i}{a(t)},
\eq
where  a dot denotes the derivative with respect to the cosmic time.  In terms of the integration constants, the Wronskian condition takes the form
\bq
\lb{3.12}
a_k b_k^*-a_k^*b_k=\frac{i\pi}{k}.
\eq 

It should be noted that the validity of our analytical approximations (\ref{3.9}) is justified only when the discarded term $\psi$ is relatively small compared with $\xi$. Our simulations show this is indeed the case if $k\ge1$. In Fig. \ref{fig3}, we choose $k=5$ and compare the analytical solution with the numerical one.  As $\psi$ is less than one percentage of $\xi$ in the bouncing and transition phases, our analytical approximations are in a good match with the numeric one.
\begin{figure}
{
\includegraphics[width=8cm]{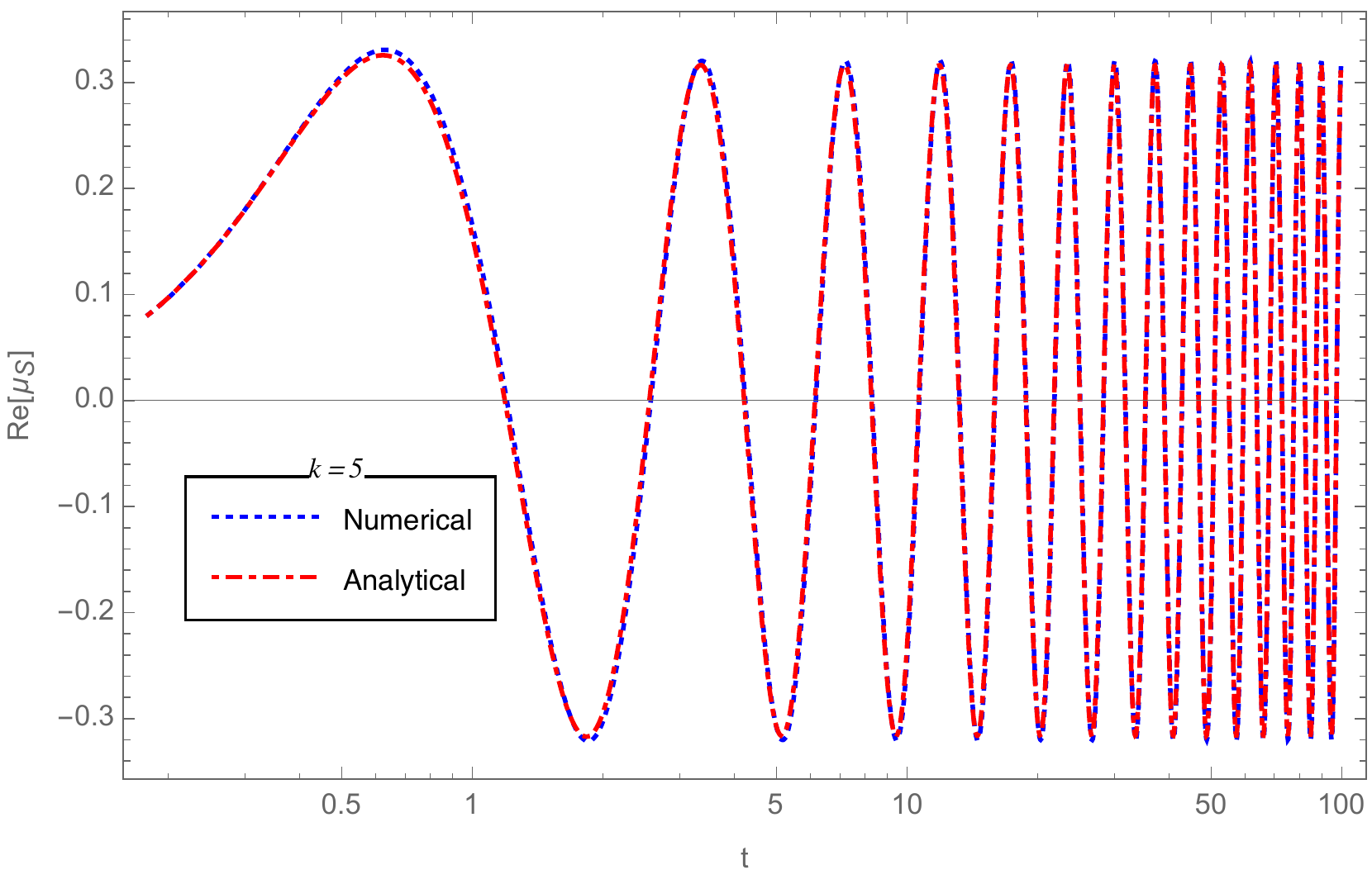}
\includegraphics[width=8cm]{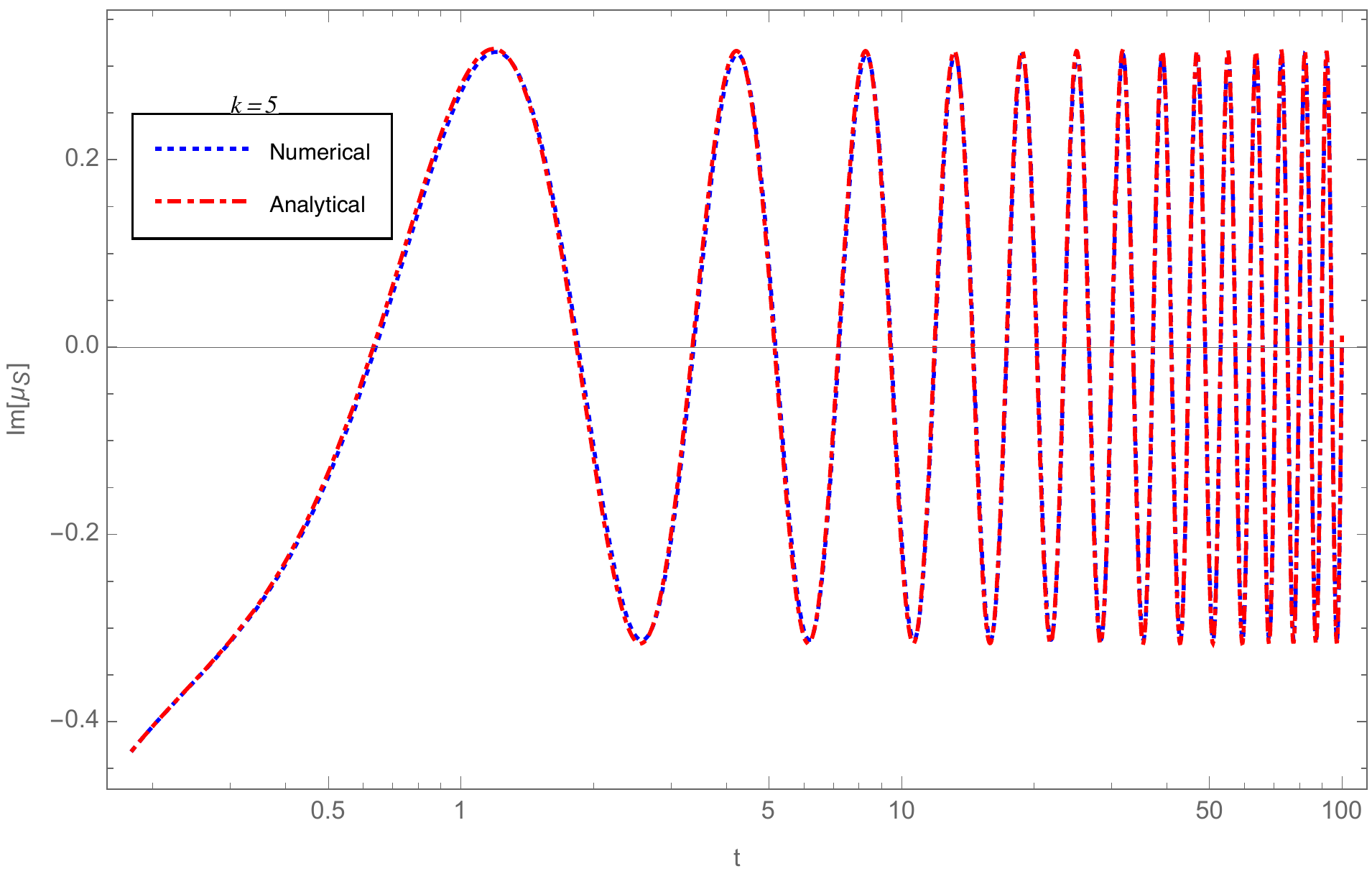}
\includegraphics[width=8cm]{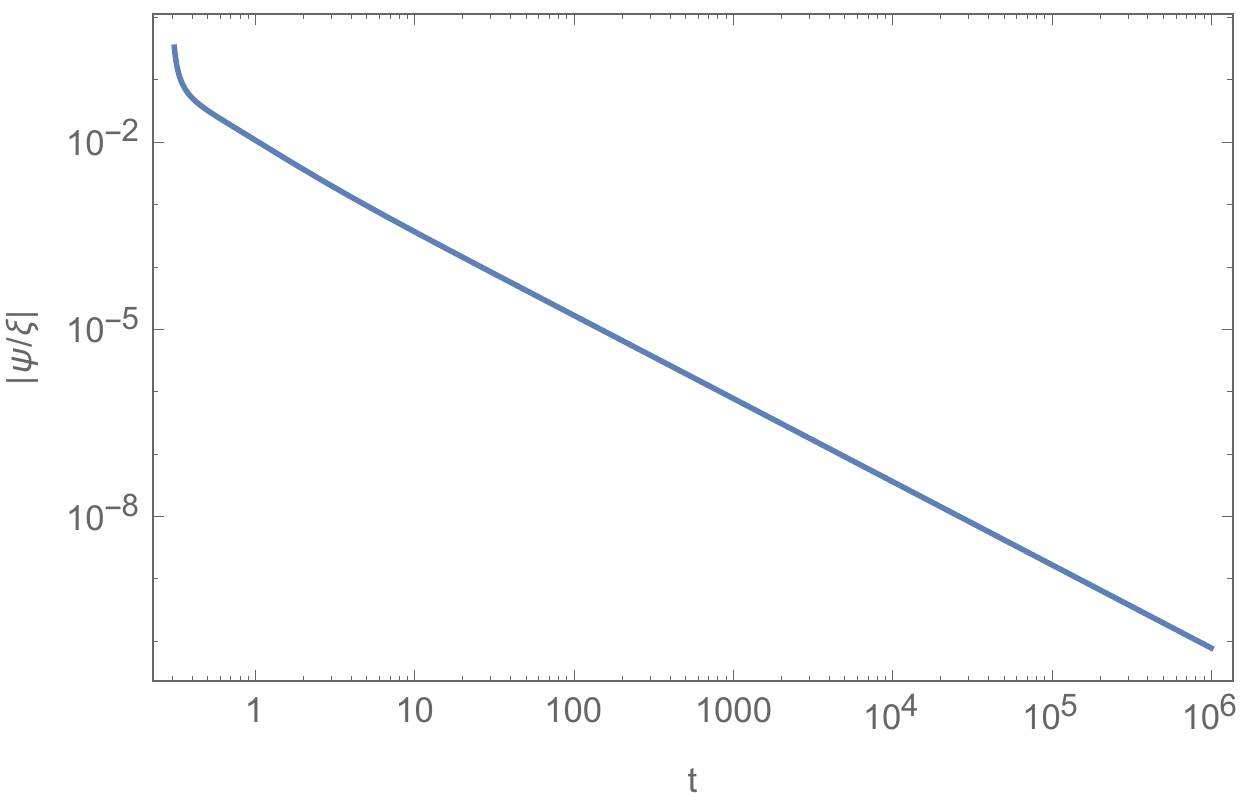}
}
\caption{In this figure, we compare the numerical solution (red) of Eq. (\ref{scalarfun}) with our analytical one (blue) in Eq. (\ref{3.9}) for $k=5$. Initial conditions are imposed at the silent point by setting $a_k=i b_k=\sqrt{\frac{\pi}{2k}}$. The first two subfigures show the difference in the real and imaginary parts of  analytical and numerical solutions. In the last subfigure, we show the comparative magnitude between $\psi$ and $\xi$ which explains why our approximations are in a good fit with the numeric solution.}
\label{fig3}
\end{figure}

\subsection{Primordial Power Spectrum in the UV Regime}

The observable window of wave numbers in the current CMB observations ranges between $k_{\text{min}}=10^{-4} \text{Mpc}^{-1}$ and $k_{\text{max}}=1 \text{Mpc}^{-1}$.\footnote{$\text{Mpc}^{-1}\approx 5.24\times 10^{-58}$ in the Planck units.}
 In LQC, at the bounce, there is a characteristic energy scale given by 
\bq
k_B\equiv a_B\sqrt{\rho_c}M^{-1}_{\text{Pl}}.
\eq  
The modes with $k\gg k_B$ are located in the ultraviolet (UV) regime while modes with $k\ll k_B$ are in the infrared (IR) regime,  and modes with $k\approx k_B$ are in the intermediate regime. As the physical lengths of the primordial perturbations are stretched out in an expanding Universe, it's important to determine which   {range of } the observable window at  {the} present  { corresponds to at the bounce}. Generally, the answer to this  question depends on the history of the Universe since
\bq
k^B_{\text{phy}}=e^N k_{\text{phy}},
\eq
where $k^B_{\text{phy}}$ represents the physical length of the mode at the bounce and $k_{\text{phy}}$ is the physical length at  {the} present, N denotes the total e-foldings from the bounce to the present. Previous investigations \cite{zwcks2017, lb2013} on the preinflationary dynamics in LQC indicate that the preferable e-foldings from the bounce to the end of inflation is around $N_\text{inf}=140$ which gives a rough estimate on the total e-foldings $N\approx 200$.  This implies that the modes with the longest wavelength that can be observed today was located in the deep UV regime at the bounce,  { $k^B_{\text{min}}\sim10^{25}$}. Of course, by fine-tuning the initial conditions in the contracting phase (or at the bounce), the observable window can also be moved to the IR regime. The minimal requirement is $k^B_\text{max} \approx 1$ which implies the e-foldings from the bounce to the end of inflation should be around   $N_\text{inf}\approx  70$.

In the UV regime, $k\gg 1$, the analytical approximations  of the turning point, scalar mode function and primordial  power spectrum can be derived order by order  in terms of the Taylor expansion of the  inverse wavenumber.   {Specially, when} near the silent point, the turning point can be expanded into the series 
\bq
\lb{tp}
t^{\text{UV}}_+ =  t_S+ \frac{\delta^{\text{UV}}_S}{k^2}+\mathcal O\left(\frac{1}{k^3}\right),
\eq
where $t_S=\frac{1}{\sqrt{\gamma_B}}$ and $\delta^{\text{UV}}_S=\sqrt{\gamma_B}\left(\frac{1}{4}+\frac{19\times 2^{1/3}}{18}\right)$. 
Consequently, 
\bq
\xi_S(t_S)\approx \left(\frac{3\sqrt{g(t_S)}\delta^{\text{UV}}_S}{2k a(t_S)}\right)^{2/3} = \xi^{\text{UV}}_S k^{-\frac{4}{3}},
\eq
where $\xi^{\text{UV}}_S=\frac{\left(9+38\times 2^{{1}/{3}}\right)\gamma^{{2}/{3}}}{12\times 2^{{7}/{9}}\times3^{{1}/{3}}}\approx 0.735$. In the above approximation, only the leading-order term is retained. Now, it's straightforward to compute the power spectrum at the silent point,  which  is
\bqn
\Delta^2_\mathcal R(t_S)&=&\lim_{\tau\to 1}\frac{k^3}{2\pi^2} \left |\frac{\mu_S}{z_S}\right |^2\nb\\
&\approx&\Delta^{\text{UV}}_Sk^{\frac{10}{3}} \Big\{  \left |a_k\right |^2 A_S^2+  \left |b_k\right |^2 B_S^2  \nb\\
&+&  a^*_k b_k  A_S B_S  +a_k b^*_k A_SB_S \Big \},
\eqn
where $\Delta^{\text{UV}}_S \equiv \frac{3^{{1}/{3}}  \gamma^{{5}/{6}}_B }{72\times 2^{{13}/{18}}\pi^2 \rho_c}$ and $B^2_S=3 A^2_S=3^{-1/3}\Gamma^{-2}(2/3)$.

In the transition phase when $t\approx 10^4\sim 10^5 $,  $\xi_S(t)$  {approaches asymptotically to  negative infinity, and} in this asymptotic region, owing to  the asymptotic forms of the Airy functions,
\bqn
\lb{asym2}
{\rm Ai} (\xi) \to \frac{1}{\sqrt{\pi} (-\xi)^{1/4}} \cos\left[\frac{2}{3}(-\xi)^{3/2}-\frac{\pi}{4}\right],\\
{\rm Bi} (\xi) \to - \frac{1}{\sqrt{\pi} (-\xi)^{1/4}} \sin\left[\frac{2}{3}(-\xi)^{3/2}-\frac{\pi}{4}\right].
\eqn
 {Then, the} mode function takes the form 
\bqn
\lb{3.13}
\mu_S&=& \frac{1}{\sqrt{\pi}(- g)^{1/4}} \left\{a_k  \cos \left[\frac{2}{3}(-\xi)^{3/2}-\frac{\pi}{4}\right] \right. \nb\\
&& \left.+ b_k \sin\left[\frac{2}{3}(-\xi)^{3/2}-\frac{\pi}{4}\right] \right\}.
\eqn
On the other hand, in the transition phase, as the Hubble horizon approaches  {to the } negative infinity as shown in Fig. \ref{fig1}, the mode  {function} equation (\ref{scalarfun}) simply reduces  to
\bq
\mu^{''}_S+k^2 \mu_S=0,
\eq
which  {has the solution} 
\bq
\lb{3.14}
\mu_S=\frac{1}{\sqrt{2k}}\left( \tilde \alpha_k e^{-i k\eta}+\tilde \beta_k e^{ik \eta}\right),
\eq
where $\tilde \alpha_k$ and $\tilde \beta_k $ are two parameters whose explicit expressions can be found from Eq.  (\ref{3.13}) in the following way. 
In the $\text{UAA}$, by definition, 
\bqn
\frac{2}{3}\left(-\xi\right)^{3/2} = k \int_{\eta_+}^\eta \sqrt{- g_S(\eta)} d\eta.
\eqn
Noting that $g_S(\eta)=g_S(t)$ quickly converges to   {the} negative unity  as depicted in Fig. \ref{fig2}, we can therefore assume that after $\eta=\eta_f$, the function $g_S(\eta) \simeq -1$. Therefore, 
\bqn
\frac{2}{3}\left(-\xi\right)^{3/2} &=& k \int_{\eta_+}^\eta \sqrt{- g_S(\eta)} d\eta\nb\\
&=& k\eta - k \eta_f + k \int_{\eta_+}^{\eta_f} \sqrt{- g_S(\eta)} d\eta.
\eqn
Defining
\bqn
\eta_{f B}& =& \eta_f -  \int_{\eta_+}^{\eta_f} \sqrt{- g_S(\eta)} d\eta,  \nb\\  
&=&\eta_B + \hat \eta (t_f),
\eqn
where $\hat \eta(t)$ is defined by
\bq 
 \hat \eta(t)= t  \,_2F_1\left(\frac{1}{6},\frac{1}{2},\frac{3}{2},- \frac{t^2}{\tau_B^2}\right)-\int^{t}_{t_+} \frac{\sqrt{-g_S(t)}dt}{a(t)}.
\eq

Now,  Eq. (\ref{3.13}) can be written into   
\bqn
\lb{3.15}
\mu_S&=& \frac{e^{-i\pi/4}}{2\sqrt{\pi}}\left\{ \left(a_k-i b_k \right) e^{ik(\eta-\eta_{f B})} \right.\nb\\
&&\left. +\left(i a_k- b_k \right) e^{-ik(\eta-\eta_{f B})}\right\}, 
\eqn
which, once compared with Eq. (\ref{3.14}), leads to the connections between the two sets of integration constants,   {given by} 
\bqn
\lb{connection1}
\tilde \alpha_k&=&\sqrt{\frac{k}{2\pi}}\left(i a_k -b_k\right) e^{i k \eta_{f B}-i\pi/4 }, \\
\lb{connection2}
\tilde \beta_k&=&\sqrt{\frac{k}{2\pi}}\left(a_k -i b_k\right) e^{-i k \eta_{f B}-i\pi/4 },
\eqn
with the Wronskian condition $|\tilde \alpha|^2-|\tilde \beta|^2=1$.

Right after the transition phase, when the energy density drops to about $10^{-12}\rho_c$, the slow-roll inflation takes place. During this phase, the mode functions can be expanded in terms of the Hankel functions. An analysis on the mode function \cite{zwcks2017} reveals that the power spectrum in this phase can be computed by the formula
\bq
\lb{power1}
\Delta^2_{\mathcal R}=| \alpha_k+ \beta_k|^2 \Delta^{{\mathcal GR}}_{\mathcal R},
\eq
where $\alpha_k=\tilde \alpha_k$, $\beta_k=\tilde \beta_k$ and the power spectrum in the classical theory  $ \Delta^{\mathcal {GR}}_{\mathcal R}$ is given by 
\bq
\Delta^{\mathcal {GR}}_{\mathcal R}\equiv \frac{k^2}{2\pi^3}\left(\frac{H}{a \dot \phi}\right)^2\Gamma^2(\nu_s)\left(\frac{-k\eta}{2}\right)^{1-2\nu_s},
\eq
with $\nu^2_s\approx \eta^2 z^{''}_S/z_S+1/4$.
In general, the power spectrum for a particular mode is computed at the horizon crossing when $k_*=a(t_*)H(t_*)$ as all the modes are frozen outside the horizon. Besides, in Eq. (\ref{power1}), the corrections due to the quantum  {gravitational}  effects near the bounce are parametrized by the coefficients $\alpha_k$ and $\beta_k$ which in turn  can be directly related  {to} the initial conditions at the silent point,  if Eqs. (\ref{connection1})-(\ref{connection2}) are taken into account.  It can be shown easily that the averaged power spectrum is related  {to} the the classical result via 
\bqn
\lb{quantumcorrections}
|\alpha_k+ \beta_k|^2&=&\frac{k}{\pi}\left( |a_k|^2+|b_k|^2\right). 
\eqn
Therefore, if we require the quantum corrections are negligible, namely, $\mathcal P_{\mathcal R}= \mathcal P^{\mathcal {GR}}_{\mathcal R}$, then the condition on the initial data at the silent point  should be 
\bq
\lb{3.16}
|a_k|^2+|b_k|^2=\frac{\pi}{k}.
\eq
The only initial data that satisfy both Eq. (\ref{3.16}) and the Wronskian condition (\ref{3.12}) is 
\bq
\lb{classicallimit}
a_k=\sqrt{\frac{\pi}{2k}}, \quad \quad b_k=- i\sqrt{\frac{\pi}{2k}},
\eq
which fix the form of initial power spectrum at the silent point to be 
\bq
\lb{3.17}
\Delta^2_\mathcal R \propto k^{\frac{7}{3}}.
\eq
Moreover, if the initial condition (\ref{classicallimit}) is substituted back into Eq. (\ref{3.15}), one can immediately find that  during the transition phase, the Universe is in the BD vacuum. This explains why (\ref{classicallimit}) can lead to the classical results.  

To confirm this point, the numeric simulations of the power spectrum are presented in Fig. \ref{scalarpower}.  In the top panel,  the initial condition (\ref{classicallimit}) at the silent surface are used. It can be seen from the graph that although the power spectrum is oscillating at large wavenumbers ($k>1$), its average value is scale-invariant and  its order of magnitude is also as expected ($\approx 2.2\times 10^{-9}$) from 2015 Planck data. We also try   several different  sets of initial conditions, and all give a divergent behavior, as shown by  the bottom panel of Fig. \ref{scalarpower}, which is consistent with the numerical  results obtained previously \cite{ bbgs2016,mbg2018}. Note that in plotting the bottom panel of Fig. \ref{scalarpower}, we have choosen
\bq
\lb{3.17a}
a_k=\frac{\pi}{2k^2}, \quad b_k=-i k,
\eq
 at the silent surface.  
 
\begin{figure}
{
\includegraphics[width=8cm]{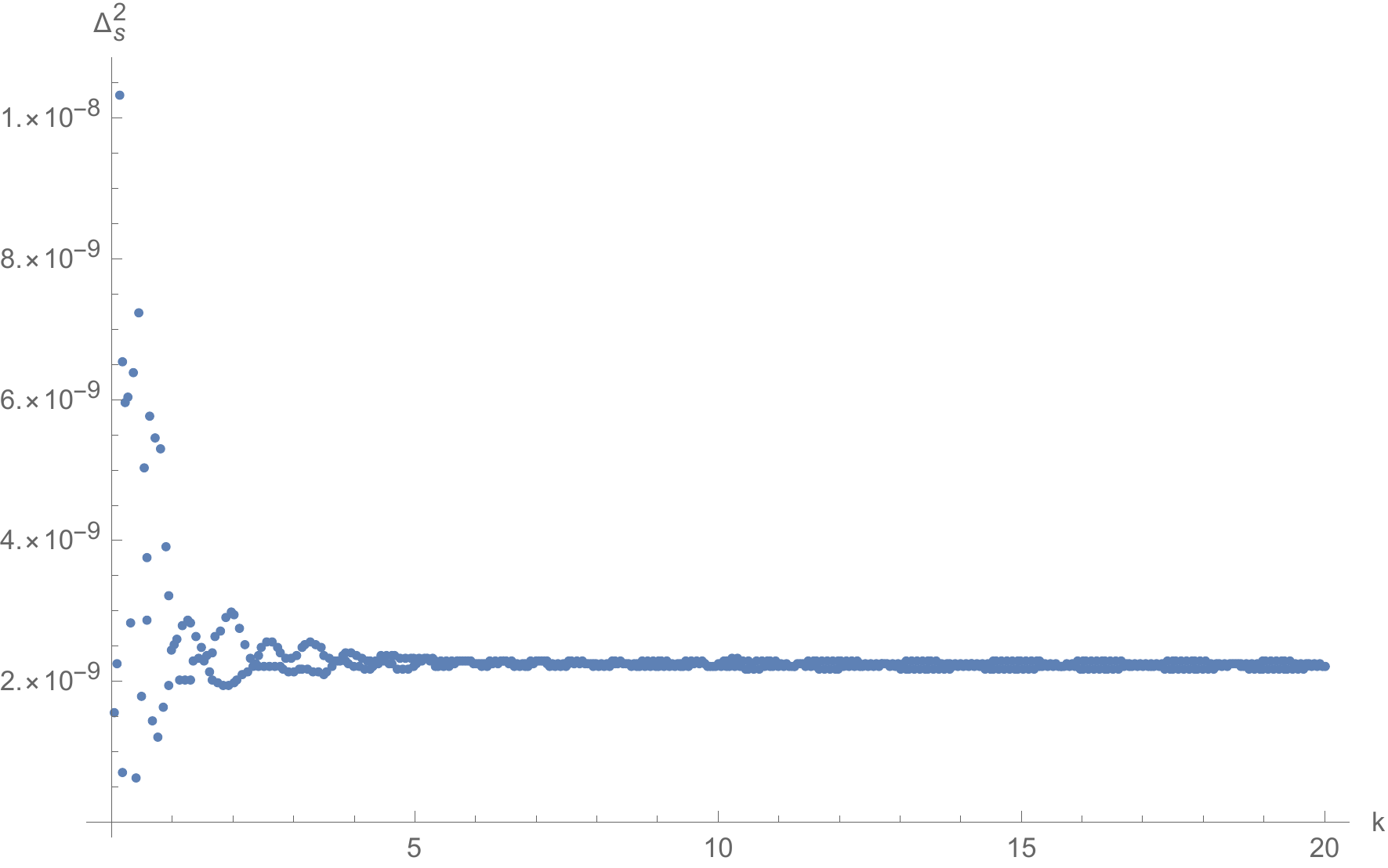}
\includegraphics[width=8cm]{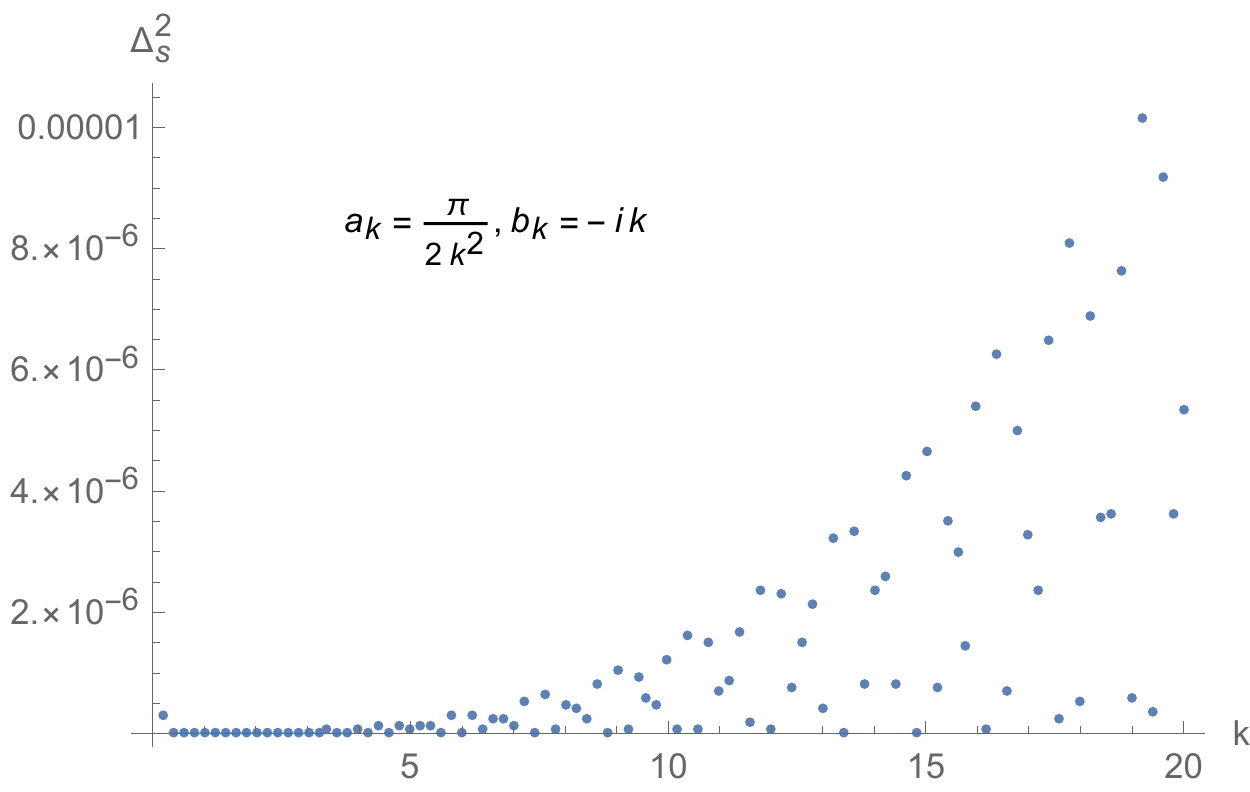}
}
\caption{Primordial power spectrum for the scalar perturbations when the Universe is filled with a single scalar field with the chaotic potential. The mass of the scalar field is chosen to be $1.26\times 10^{-6}$ and the initial condition for the background evolution is set at the bounce with $\phi_B=5.6$ and $\dot \phi<0$. The top panel is the power spectrum resulting from the initial condition (\ref{classicallimit}) which leads to a scale-invariant power spectrum at large wavenumbers,  while the bottom panel corresponds to the initial condition(\ref{3.17a}).}
\label{scalarpower}
\end{figure}

To conclude this section, we would like to show that the initial condition (\ref{classicallimit}) is in fact  the only one that will generate the primordial power spectrum equal to the classical result in the observable regime,  {and} other choices of the initial conditions will inevitably lead to a correction term proportional to some positive power of the co-moving wavenumber, and  {hence} a divergent power spectrum in  {the} UV regime.  In order to prove this, one can start with a more general parameterization
\bq
\lb{3.18}
a_k= a_0 k^n, \quad \quad b_k= b_0 e^{-i\theta}k^l,
\eq
where $\theta$ is the relative phase between $a_k$ and $b_k$.  Besides, both $a_0$ and $b_0$ are positive and independent of $k$. With these new parameters, the Wronskian condition is now equivalent to 
\bq
\lb{3.19}
2a_0b_0\sin \theta=\pi, \quad \quad n+l=-1.
\eq
 {Thus, the quantum corrections can now be written as}
\bq
\lb{3.20}
|\alpha_k+ \beta_k|^2=\frac{1}{\pi}\left( a^2_0 k^{2n+1}+b^2_0 k^{-2n-1}\right).
\eq
Combining Eqs. (\ref{3.19}) and (\ref{3.20}), one can immediately conclude 
\bq
\lb{3.21}
|\alpha_k+ \beta_k|^2=1 \iff  n=l=-\frac{1}{2},\;\;\;  \sin \theta =1.
\eq

\subsection{Primordial Power Spectrum in the IR Regime}

The calculations of the power spectrum in the IR regime has already been discussed in great details in \cite{bgsl2015}. Although, in their paper, the initial conditions are chosen in the remote past of the contracting phase, the results are actually independent of the time  {(up to a rescaling of the power spectrum), once} the initial conditions are chosen. The reason is as follows. In the IR regime as $k\ll1$, throughout its evolution in the bouncing and transition phases, the mode  {function} equation (\ref{scalarfun}) has the approximate solution 
\bq
\lb{3.22}
\mu_S=a_k z_S+b_k  z_S \int^\eta_{\eta_*} \frac{d\eta'}{z^2_S}+\mathcal O(k^2),
\eq
where $\eta_*$ denotes some particular time. As a result, the power spectrum at any time is given by 
\bqn
\lb{3.23}
\Delta^{\text{IR}}_S(\eta)&=&\frac{k^3}{2\pi^2} \left |\frac{\mu_S}{z_S}\right |^2, \nb\\
&=&\frac{k^3}{2\pi^2} \left |a_k+b_k  \int^\eta_{\eta_*} \frac{d\eta'}{z^2_S}\right |^2.
\eqn
In general, the power spectrum can be obtained by evaluating the integral in Eq. (\ref{3.23}) explicitly at the end of the slow-roll inflation $\eta=\eta_\text{end}$.  Here, the key observation is that the integral $ \int^{\eta_\text{end}}_{\eta_*} d\eta'/z^2_S$  turns out to be only dependent on the background dynamics and therefore irrespective of any particular mode. All the information concerning the wavenumber $k$ is incorporated into the integration constants $a_k$ and $b_k$ which satisfy the Wronskian condition
\bq
\lb{3.24}
a_k b^*_k-a^*_k b_k=i. 
\eq
With the parameterization (\ref{3.18}),  the above Wronskian condition is equivalent to 
\bq
\lb{3.25}
2a_0b_0\sin \theta=1, \quad \quad n+l=0.
\eq
Thus, the only set of initial conditions that can lead to a scale-invariant power spectrum is the one chosen in \cite{bgsl2015}, namely $l=-n=-3/2$,  so that 
\bq
\lb{scaleinvariant}
a_k\propto k^{3/2}, \quad \quad b_k\propto k^{-3/2}.
\eq
With such a choice, the power spectrum turns out to be scale-invariant at any given time from the bouncing phase to the end of slow-roll inflation.

\section{The Primordial Tensor Perturbations}
\renewcommand{\theequation}{4.\arabic{equation}}\setcounter{equation}{0}

The analysis of tensor perturbations in closed algebra approach can be carried out in the same  {way} as the scalar perturbations  {given in the last section}. The only complications come from the fact that the tensor mode function becomes divergent at the silent point. However, with the analytical approximations from UAA, we can easily regularize the divergence. Finally, a finite power spectrum at the silent point is found and an explicit relation between the initial conditions at the silent point and  the resulting power spectrum  is  {given}. 

\subsection{The Mode  Function and  Co-moving Hubble Horizon}

The equation of motion for the primordial tensor perturbations  takes the form \cite{bhks2008, cmbg2012, cbgv2012}
\bqn
\lb{4.1}
\mu_T''(\eta) + \left[\Omega(\eta)k^2- \frac{z_T''}{z_T} \right]\mu_T(\eta)=0,
\eqn
where $\Omega(\eta)$ is defined in Eq.~(\ref{omega}) and
\bqn
\lb{4.2}
z_T(\eta) = \frac{a(\eta)}{\sqrt{\Omega(\eta)}}.
\eqn
It should be noted that unlike $z_S$ in  {the} scalar perturbations,  $z_T$ is singular at the silent point, which  {leads to   the mode function singular at the silent point, too}. The background evolution is still described by Eqs. (\ref{scalar_analytical}) and (\ref{phi_sol}).
Analogous to the Hubble horizon (\ref{scalar_horizon}) in Sec. III,  we can also define the Hubble horizon of the tensor perturbations via
\bq
\lb{tensor_horizon}
\lambda^2_T=  \Omega \frac{ z_T}{z_T^{''}}.
\eq
As shown in Fig. \ref{fig4}, the qualitative behavior of the Hubble horizon in  {the} tensor perturbations is analogous to  {the scalar ones}. The only quantitative difference occurs at the  {bounce. In} the scalar perturbation, $\lambda^2_S(t_B)=0$ while in tensor perturbations,  $\lambda^2_T(t_B)\approx -0.014$. As a result, all the analysis  given in Sec. III. A on the Hubble horizon can also apply to the tensor case. In particular, all the modes are initially outside the horizon at the bounce  {  and later enter  the horizon at a time that is } in the interval $t_S<t<t_H$. Afterwards, these modes remain inside the horizon when $t_H<t<t_i$ and then exit the horizon during the slow-roll inflation. 

\begin{figure}
{
\includegraphics[width=8cm]{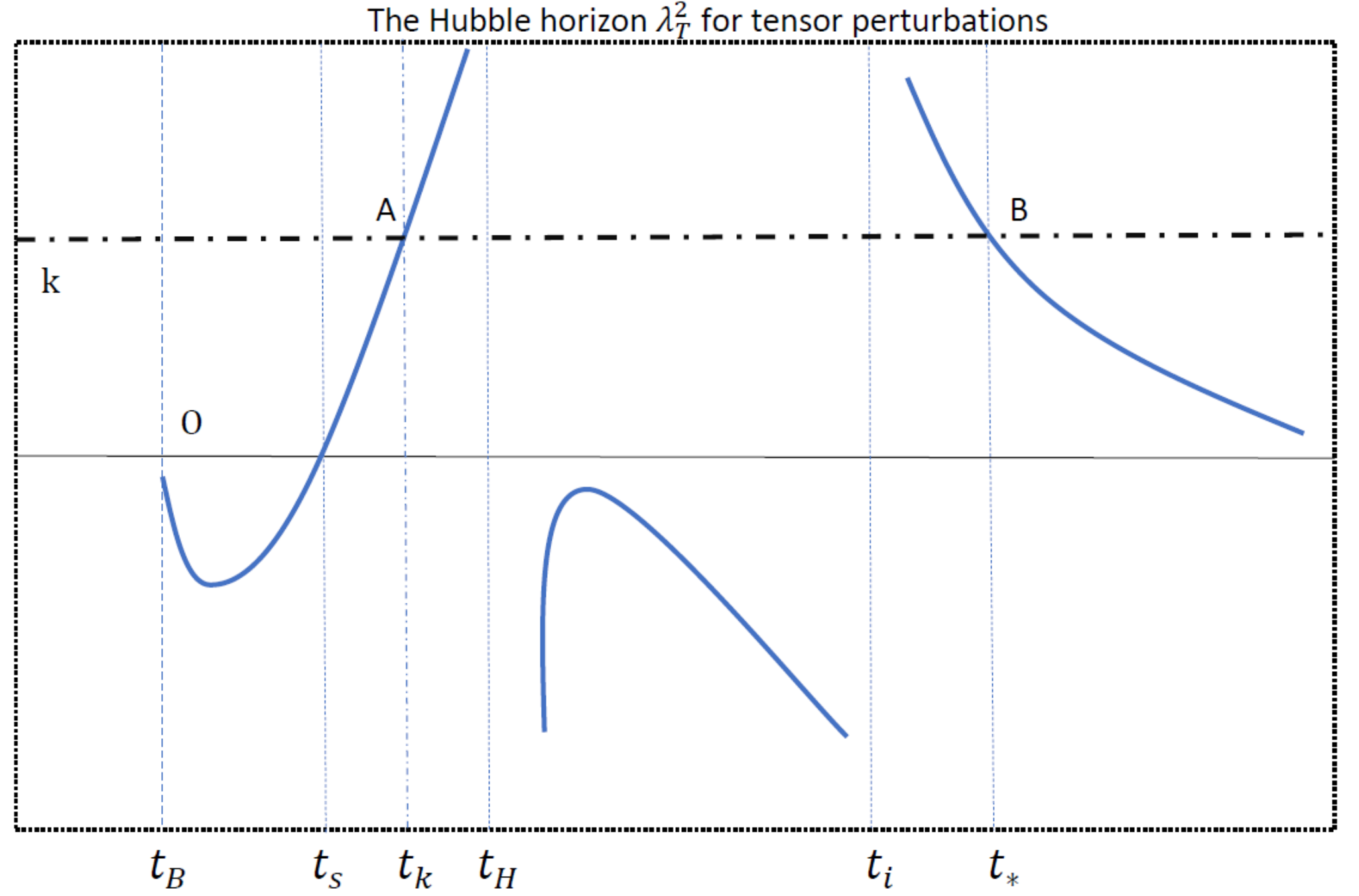}
}
\caption{A schematic plot of the Hubble horizon in the tensor perturbation. In the figure, the solid curves depict the behavior of the Hubble horizon $\lambda^2_T$ in the pre-inflationary phase. $t_B$ denotes the bounce time, $t_S$ is the time when the super-inflationary phase ends,  $t_H\approx 0.91t_{\text{Pl}}$ is the time when $z^{''}_T=0$ and $t_i$ again signifies the onset of  {the} slow-roll inflation. At the bounce, $\lambda^2_T(t_B)=-0.014$. A particular mode with co-moving wavenumber $k$ is also marked out. It enters the horizon at  point $A$ when $t=t_k$ and exits  {it}   in the inflationary phase at point $B$ when $t=t_*$.}
\label{fig4}
\end{figure}

In order to find   {analytical solutions, we first write Eq. (\ref{4.1}) in  the standard form with the  new argument $y=- k \eta$, }
\bqn
\lb{4.4}
\frac{d^2\mu_T(y)}{dy^2} = \Big\{ g_T(y) + q_T(y) \Big\} \mu_T(y),
\eqn
where
\bq
\lb{4.5}
g_T(y) + q_T(y)= \frac{\left(1-\tau^2\right)}{1+\tau^2}+\frac{\gamma_B\left(21+47\tau^2+41\tau^4-\tau^6\right)}{9k^2\left(\tau^2-1\right)^2\left(\tau^2+1\right)^{5/3}},
\eq
which has a  pole located at the silent point $\tau^2=1$ since
\bq
\lb{4.6}
g_T(y) + q_T(y)  \to  \frac{^3\sqrt{54}  \gamma_B}{k^2\left(\tau^2-1\right)^2}.
\eq
 {Following} the analysis of  {the} error control function, for a second-order pole, one has to choose \cite{zwcks2016}
\bq
\lb{4.7}
q_T(y)  = - \frac{2^{{1}/{3}}\gamma_{\rm B}}{k^2(1- \gamma_{\rm B} t^2)^2},
\eq
which leads to
\bqn
 \lb{gofy2}
 g_T(y) &= &\frac{\left(1-\tau^2\right)}{1+\tau^2}+\frac{2^{{1}/{3}}\gamma_B}{k^2\left(\tau^2-1\right)^2}\nb\\&+&\frac{\gamma_B\left(21+47\tau^2+41\tau^4-\tau^6\right)}{9k^2\left(\tau^2-1\right)^2\left(\tau^2+1\right)^{5/3}}.
\eqn

The behavior of $g_T$ is analogous to $g_S$ presented in Sec. II.B, that is, $g_T$ has only one single turning point in the bouncing and transition phases and also quickly converges to  { the negative unity}. Consequently, all the formulae in Sec. II.B for the scalar modes are also applicable to the tensor modes as long as $g_S$ is replaced by $g_T$ given in Eq. (\ref{gofy2}). In particular, the analytical solution of  {the} tensor modes is given by 
\bq
\lb{4.8}
\mu_T = \left(\frac{\xi_T}{g_T}\right)^{1/4} \Big\{a_k \rm{Ai}\left(\xi_T\right) +b_k\rm{Bi}\left(\xi_T\right)\Big\},\\
\eq
 with  
\bqn
\lb{4.9}
\xi_T = 
\begin{cases}
 \left(- \frac{3k}{2} \int_{t_{+}}^t \frac{\sqrt{ g_T(t)}}{a(t)} dt \right)^{2/3}, &\;  t< t_+ \\
- \left( \frac{3k}{2} \int_{t_{+}}^t \frac{\sqrt{-  g_T(t)}}{a(t)} dt \right)^{2/3}, &\;  t> t_+,
\end{cases}
\eqn
where $t_+$ represents the turning point of $g_T$, i.e. $g_T(t_+)=0$.
 In Fig. \ref{fig6}, we compare the analytical  {solution}  with the numerical  {one  for} $k=10$. As the mode function becomes divergent at the silent point, the initial conditions are chosen at $t=5$ with $a_k=i b_k=\sqrt{\frac{\pi}{2k}}$. The smallness of the ratio $\psi/\xi_T$ in the last subfigure indicates Eq. (\ref{4.8}) is a good approximation in the bouncing and transition phases. We have also checked our  {solution in the UV regime for $k>1$, and found that} our analytical approximations are valid and fit with the numeric ones very well. However, in the IR regime, as $\psi$ becomes  the same order of  $\xi$, higher-order corrections need to be taken into account. 

\begin{figure}
{
\includegraphics[width=8cm]{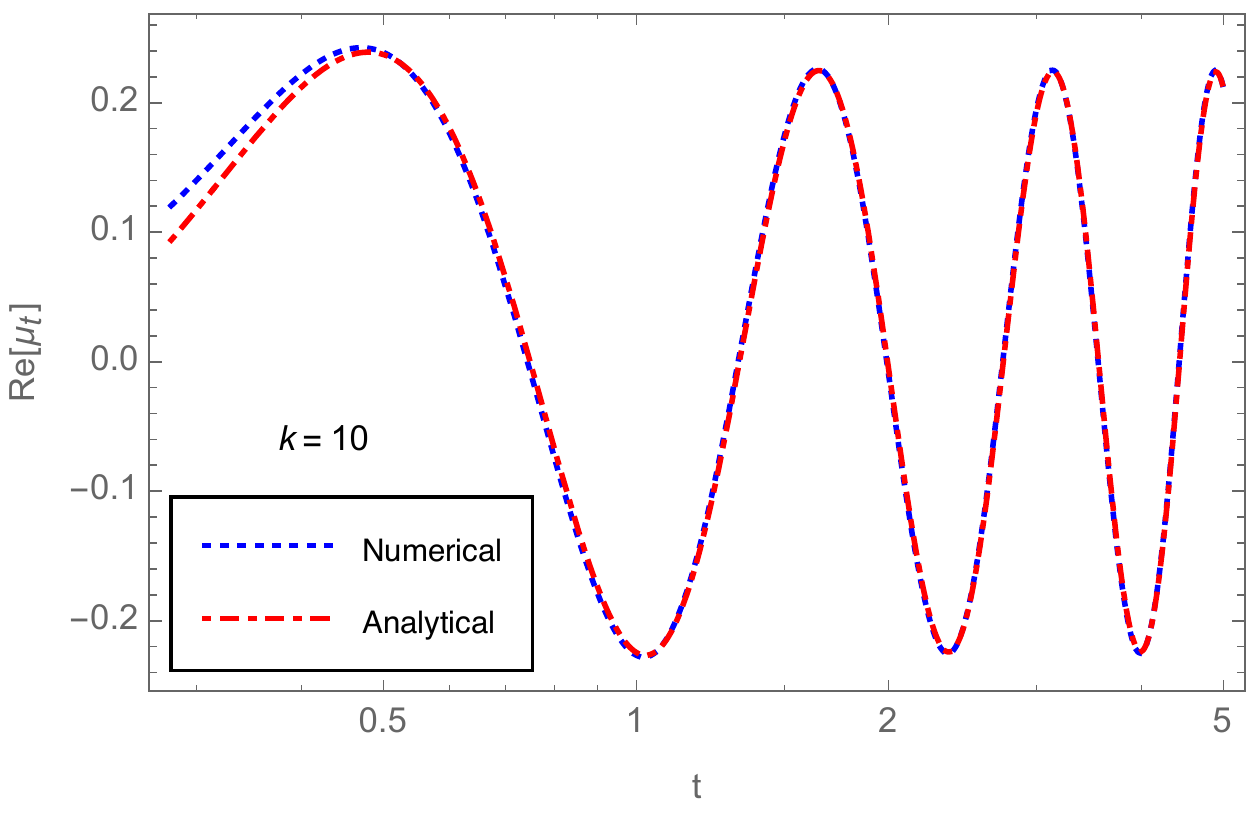}
\includegraphics[width=8cm]{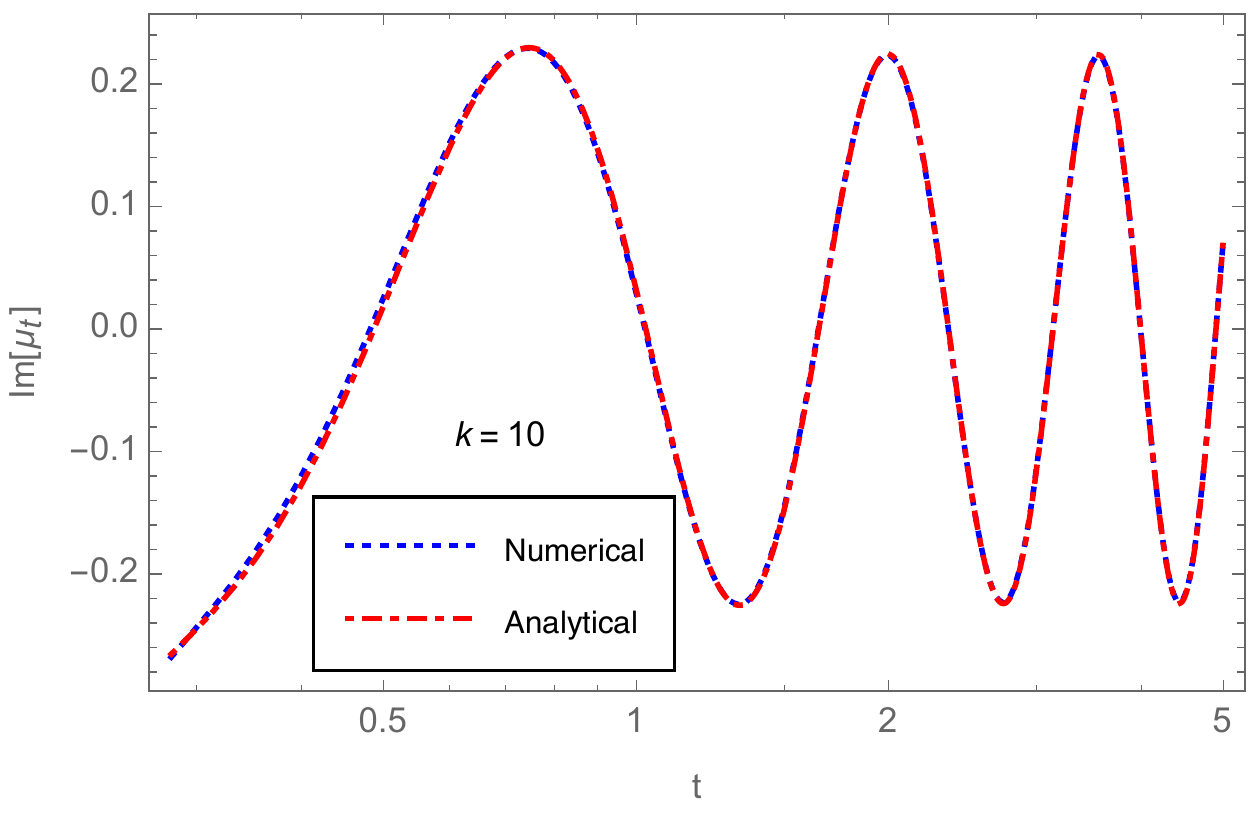}
\includegraphics[width=8cm]{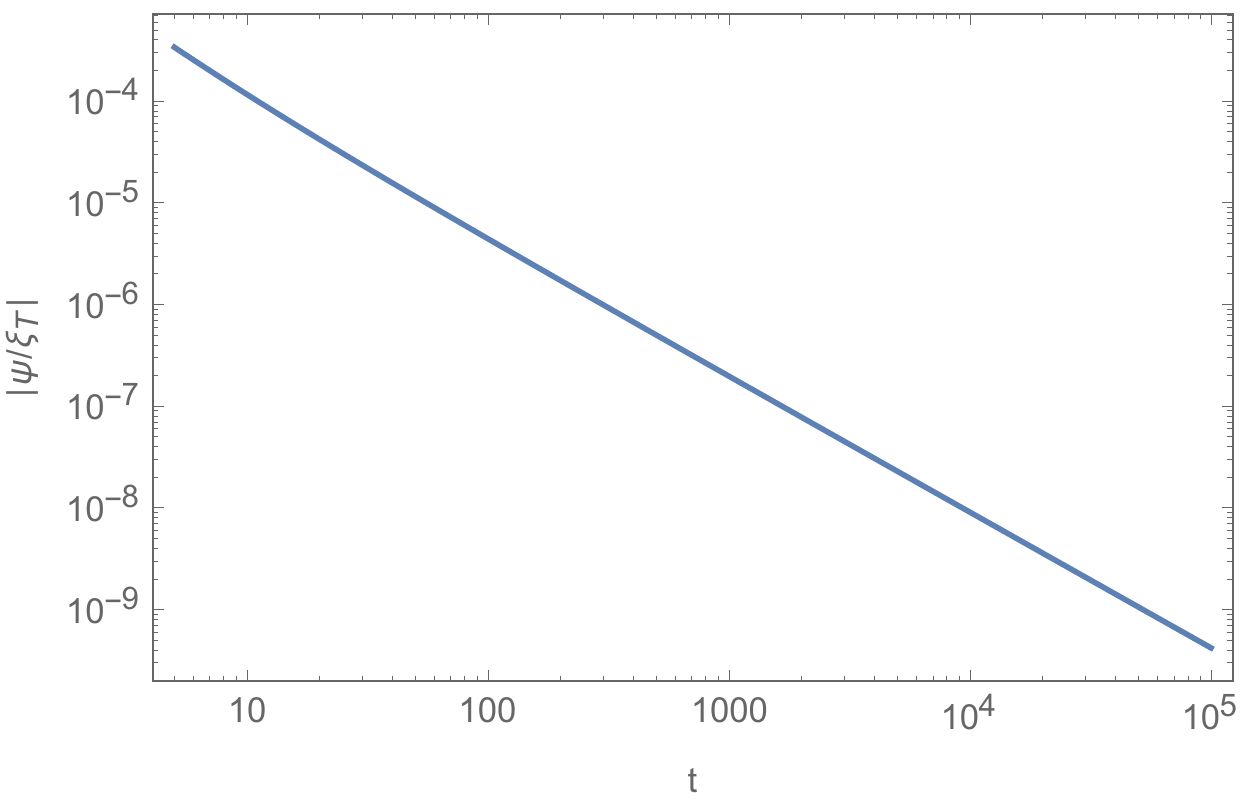}
}
\caption{In this figure, approximate solutions (\ref{4.8}) (red) are compared with the numerical  {one} (blue) by choosing the initial conditions at  {the} time $t=5$ with $a_k=i b_k=\sqrt{\pi/2k}$. In the last subfigure, the ratio $\psi/\xi_T$ is depicted until the transition phase $t=10^5$.}
\label{fig6}
\end{figure}

\subsection{Primordial Power Spectrum}  

In this section,  we  {will consider  the asymptotic forms of our analytical solutions in both of the UV and IR regimes, although we will pay more attention in the UV regime.}

 In the limit $k\gg1$,  the turning point $t_+$ has the following expansion
\bq
\lb{4.11}
t_+=\frac{1}{\sqrt{\gamma_B}}+\frac{\delta_1}{k^{2/3}}+\mathcal O\left(\frac{1}{k^{4/3}}\right),
\eq 
which, once plugged into Eq. (\ref{gofy2}), yields $\delta_1=\frac{2^{1/9}}{\gamma^{1/6}_B}\approx 0.61$.  As $\delta_1$ is a positive number, the relation $t_+>t_S$ always  holds in the UV regime. Therefore, only the upper  branch  of Eq. (\ref{4.9})  is required.  To regularize its divergence at the silent point, we can define 
\bq
\lb{4.12}
\xi_T(t_S)\equiv \lim_{\epsilon\rightarrow 0} \xi_T(t_S+\epsilon).
\eq
From the Taylor expansion of the integrand of Eq. (\ref{4.9}) 
\bq
\lb{4.13}
\frac{\sqrt{g_T(t)}}{a(t)}=\frac{1}{k (t-t_S)}-\frac{5\sqrt{\gamma_B}}{12k}+ \mathcal O\left(\epsilon\right),
\eq
 {we find that} $\xi_T$ around the silent point can be shown as 
\bq
\lb{4.14}
\xi_T(t_S+\epsilon)=\left [\frac{3}{2}\ln\left(\frac{\delta_1}{k^{2/3}}\right)-\frac{3 }{2} \ln \epsilon+ \mathcal O (\epsilon)\right ]^{2/3}.
\eq
Thus, it becomes obvious that as $\epsilon$ tends to zero, $\xi_T$ goes to positive infinity and consequently the mode function  becomes divergent at the silent point. By virtue of the asymptotic forms of the Airy functions, 
\bqn
\lb{asym3}
{\rm Ai} (\xi_T) \to\frac{e^{-\frac{2}{3}\xi_T^{3/2}}}{2\sqrt{\pi}\xi_T^{1/4}}, \quad 
{\rm Bi} (\xi_T) \to \frac{e^{\frac{2}{3}\xi_T^{3/2}}}{\sqrt{\pi}\xi_T^{1/4}},
\eqn
 {as $\xi_T \to +\infty$,}  it's straightforward to show that
\bqn
\lb{4.15}
\mu_T(t_S)&=&\lim_{\epsilon \to 0} \frac{b_k e^{\frac{2}{3}\xi_T^{3/2}}}{g^{1/4}_T \sqrt{\pi}},\nb\\ 
&=&\lim_{\epsilon \to 0} \frac{b_k \delta_2 k^{-1/6}\epsilon^{-1/2}}{\sqrt{\pi}},
\eqn
with $\delta_2=\left(\frac{2}{\gamma^6_B}\right)^{\frac{1}{36}}$. Now the power spectrum at the silent point can be computed as 
\bqn
\lb{4.16}
\Delta^2_t(t_S) &\equiv& \lim_{t \to t_S} \frac{32 k^3}{\pi}\left|\frac{\mu_T}{z_T}\right|^2,\nb\\
&=&\frac{2^{\frac{83}{18}}|b_k|^2 \gamma_B^{\frac{1}{6}}k^{\frac{8}{3}}}{\pi^2}.
\eqn
It's interesting to observe that even though our analytical mode function (\ref{4.15}) becomes divergent at the silent point, the resulting power spectrum (\ref{4.16}) is still finite. Besides, it  only depends on the magnitude of parameter $b_k$ of the mode function. The effects of $a_k$ simply dies aways as $t \to t_S$. Following the same line of arguments in Sec. II, if we require  that  the quantum corrections  {be negligible},  the only possible initial conditions at the silent point are  given by Eq. (\ref{classicallimit}) which in turns fix the initial power spectrum to be 
\bq
\lb{tensorps}
\Delta^2_t(t_S) \propto k^{\frac{5}{3}}.
\eq
All the other choices of the initial conditions will result in power spectra that become divergent in the UV limit, and hence are inconsistent with observations. In Fig. \ref{tensorpower}, the numeric simulations of the power spectrum for the tensor modes are presented with the initial conditions (\ref{classicallimit}) (top panel) and $\left(a_k=\frac{\pi}{2k^2}, b_k=-i k\right)$ (bottom panel). With the initial condition (\ref{classicallimit}),  the averaged power spectrum at large wavenumber become scale-invariant and its magnitude is around $3.3\times10^{-10}$,  while the numeric simulations with $a_k$ and $ b_k$ given by Eq.(\ref{3.17a}) at the silent surface display a divergent behavior of the power spectrum at large $k$. As a matter of fact, the initial condition (\ref{classicallimit}) is the only one that leads to a power spectrum that is consistent with current observations. The proof is quite similar to the scalar case, and we shall not repeat it here. 

\begin{figure}
{
\includegraphics[width=8cm]{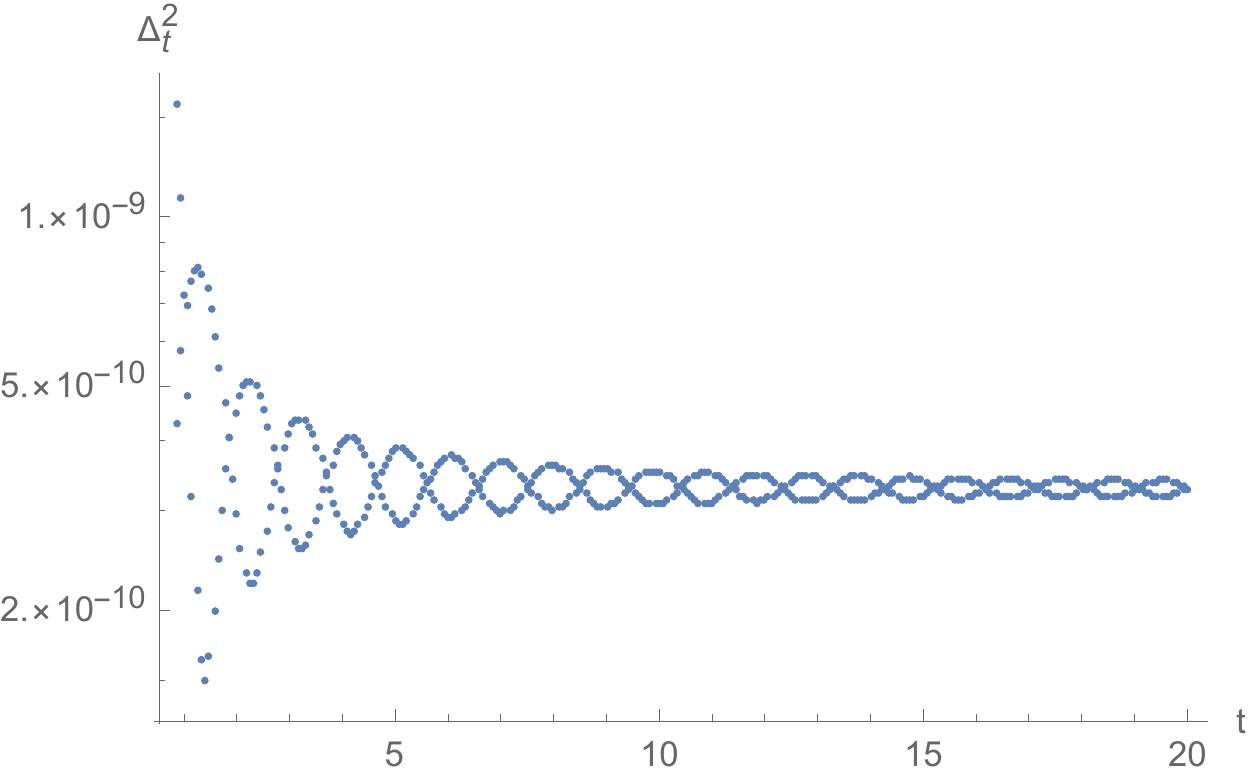}
\includegraphics[width=8cm]{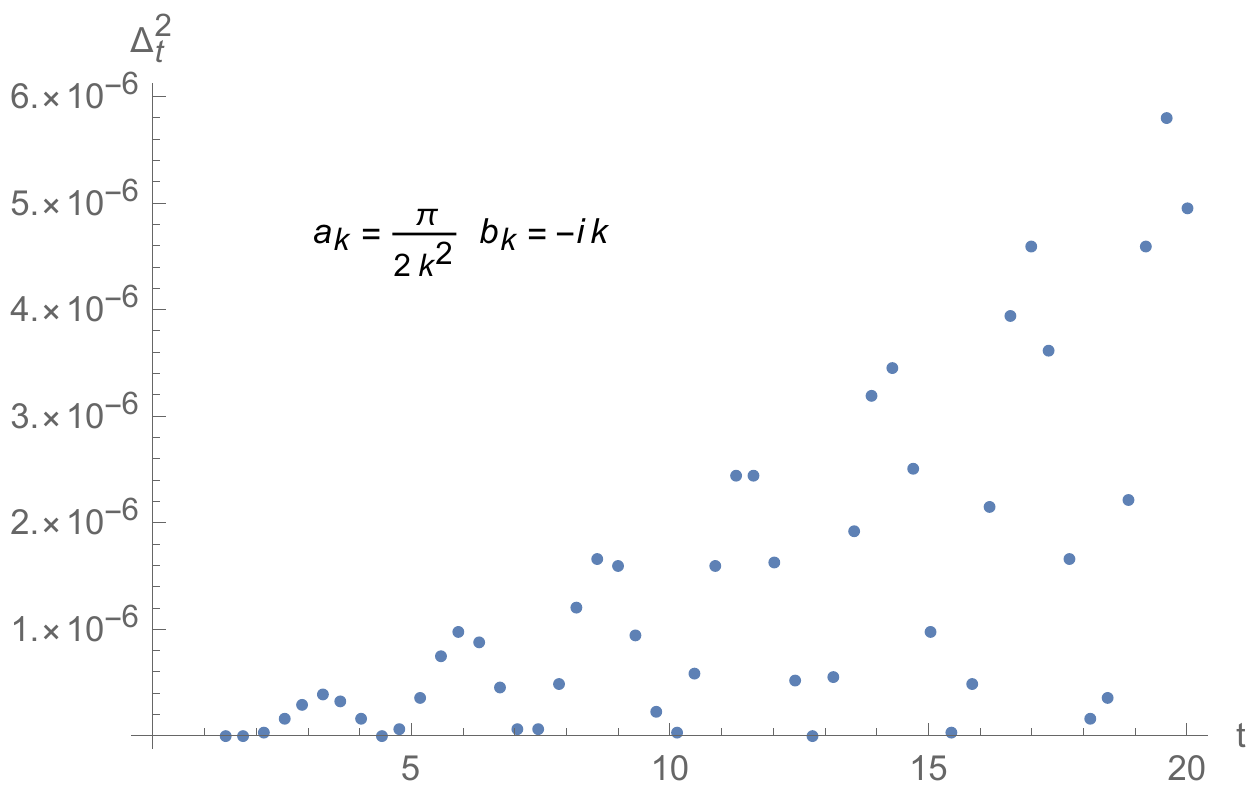}
}
\caption{Primordial power spectrum for the tensor modes when the Universe is filled with a single scalar field with the chaotic potential. The mass of the scalar field and the initial conditions for the background evolution are the same as in Fig. \ref{scalarpower}. The top panel is the result from the initial condition (\ref{classicallimit}),  while in  the bottom panel, the initial conditions are set to  (\ref{3.17a}) at the silent surface.}
\label{tensorpower}
\end{figure}
 
The calculations of the tensor power spectrum in the IR regime proceeds in the same way as for the scalar perturbations. All the formulae in Sec. III. D are still valid  {here} as long as $z_S$ in Eqs. (\ref{3.22}) and (\ref{3.23}) is replaced by $z_T$.
 {Then, it can be shown that, in  order to produce} a scale-invariant power spectrum in the IR regime, the only possible choice of $n$ and $l$ is still  $l=-n=3/2$.  {Such resulted  power spectrum in fact is  scale-invariant not only
 at the silent point but  also at any} given time in the bouncing, transition and inflationary phases.

\section{Summary}

In this work,  { using the uniform asymptotic approximations (UAA), we have explicitly derived   the analytical solutions} of both scalar and tensor mode functions in the closed algebra approach in LQC.  We have also discussed the co-moving Hubble horizon when the bounce is dominated by the kinetic energy of the scalar field and found all the modes are outside the Hubble horizon at the silent point ($ \approx0.18 t_{\text{Pl}}$). Shortly after the super-inflationary phase, all these modes enter 
 { the} Hubble horizon at  { about} $t_{\text{Pl}}$  and then re-exit the horizon during the slow-roll inflationary phase. In order to avoid the problem  { caused by the } signature change in the super-inflationary phase, our initial conditions are chosen  { to be imposed} at the silent point. Matching the solutions in the bouncing, transition and inflationary phases, we have shown how the quantum corrections are related  {to} the initial conditions at the silent point  for  both scalar and tensor perturbations. 

Although previous numerical simulations show a divergent behavior of the power spectra in the UV regime, we  have found  a particular set of initial conditions at the silent point  {[cf. Eq. (\ref{classicallimit})],  which can reproduce  results that are consistent with current observations}.  In doing so, it turns out that the initial power spectra also depend on the wavenumber as shown in Eqs. (\ref{3.17}) and (\ref{tensorps}) for the scalar and tensor perturbations, respectively. All the other choices of the initial conditions (at the silent point) would inevitably lead to divergent power spectra in the UV regime.  On the other hand, the calculations in the IR regime are based on the fact that there exist analytical solutions of the mode  function equations when the term $\Omega k^2$ is ignored. It turns out that the initial conditions at the silent point completely determines the k-dependence of the power spectra in the IR regime. More specifically, the scale-invariant properties of the power spectra in the IR regime demands the conditions (\ref{scaleinvariant}) to be satisfied. This choice is equivalent to the Bunch-Davies vacuum  in the contracting phase and it will result in scale-invariant power spectra at any given time in the bouncing, transition and inflationary phases.

 Finally, we would like to emphasize that in this work, we have applied UAA to the equations of motion for cosmological perturbations originally proposed in \cite{cbgv2012}  and assumed that they are valid for all modes.
However, as argued in \cite{we2017},  the deformed  algebra approach is valid only in the  {regime} where the dynamics can be well approximated by the effective equations. This indicates the large quantum fluctuations are ruled out in this approach from the beginning. As a result, the  closed algebra approach is only applicable to the long-wavelength modes. Then, in order to modify the UV behavior, it's essential to add quantum corrections to account for the UV modes. One of the tentative approaches is MDRs discussed in \cite{mbg2018}.   We wish to  come to this issue in another occasion.

\section{ACKNOWLEDGEMENTS}

This work is supported in part by the National Natural Science Foundation of China (NNSFC) with the
Grants Nos. 11847216, 11675143, 11675145, and 11375153.

\end{document}